\documentclass[11pt,a4paper]{article}
\usepackage{jheppub}

\newcommand{\beq}{\begin{equation}}
\newcommand{\eeq}{\end{equation}}
\newcommand{\bea}{\begin{eqnarray}}
\newcommand{\eea}{\end{eqnarray}}

\newcommand{\ZZ}{\mathbb{Z}}

\newcommand{\mc}[1]{\mathcal{#1}}

\newcommand{\ii}{\mathrm{i}}
\newcommand{\ee}{\mathrm{e}}

\newcommand{\id}{I}

\newcommand{\tr}{\mathrm{tr}}

\newcommand{\ket}[1]{|#1 \rangle}

\newcommand{\brakett}[1]{\langle #1 \rangle}

\def\preprintnumber#1#2#3{\hfill
\begin{minipage}{1.2in}
#1 \par\noindent #2 \par\noindent #3 
\end{minipage}}
\renewcommand{\thefootnote}{\fnsymbol{footnote}}

\def\comma      { \, , }
\def\period     { \, . }
\def\nn     { \nonumber }

\newcommand{\bbZ}{{\mathbb Z}}
\def\calM   {{\cal M}}
\def\ch   {{\rm ch}}
\def\calO {{\cal O}}
\def\calF {{\cal F}}
\def\del { \partial }

\title{On the mass-coupling relation of multi-scale quantum integrable models}

\author[a]{Zolt\'an Bajnok,}
\author[a]{J\'anos Balog,}
\author[b]{Katsushi Ito,}
\author[c]{Yuji Satoh}
\author[a]{and G\'abor Zsolt T\'oth}

\affiliation[a]{MTA Lend\"{u}let Holographic QFT Group, Wigner Research Centre\\
H-1525 Budapest 114, P.O.B. 49, Hungary}
\affiliation[b]{Department of Physics, Tokyo Institute of Technology\\
Tokyo 152-8551, Japan }
\affiliation[c]{Institute of Physics, University of Tsukuba\\
Ibaraki 305-8571, Japan}

\emailAdd{bajnok.zoltan@wigner.mta.hu}
\emailAdd{balog.janos@wigner.mta.hu}
\emailAdd{ito@th.phys.titech.ac.jp}
\emailAdd{ysatoh@het.ph.tsukuba.ac.jp} 
\emailAdd{toth.gabor.zsolt@wigner.mta.hu}

\abstract{We determine exactly the mass-coupling relation
for the simplest multi-scale quantum integrable model,
the homogenous sine-Gordon model with two independent mass-scales.
We first reformulate its perturbed coset CFT description in terms
of the perturbation of a projected product of  minimal models. This representation enables us to
identify conserved tensor currents on the UV side. These UV operators are then mapped via form
factor perturbation theory to operators on the IR side, which are characterized by their form factors.
The relation between the UV and IR operators is given in terms of the sought-for mass-coupling relation.
By generalizing the $\Theta$ sum rule Ward identity we are able to derive differential equations
for the mass-coupling relation, which we solve in terms of hypergeometric functions. We 
check these results against the data obtained by numerically solving the thermodynamic Bethe Ansatz 
equations, and  find a complete agreement.}

\begin{document}

\preprintnumber{TIT/HEP-653}{UTHEP-684}{}

\maketitle

\renewcommand{\thefootnote}{\arabic{footnote}}
\setcounter{footnote}{0}

\section{Introduction}
\label{sec.intr}

There have been an increasing interest and relevant progress in studying
$1+1$ dimensional integrable quantum field theories (QFTs), due to
the fact that they can be solved exactly.
The usual definitions of QFTs are based on a Lagrangian and 
the main\footnote{Some non-perturbative methods exist especially for 
supersymmetric QFTs.}
analytical tool to investigate them is perturbation theory, which
provides a systematic expansion of physical quantities around a
properly chosen free theory. In general, only a few terms are calculable
technically, leading to merely approximate results.

Integrable $1+1$ dimensional QFTs are special in the sense that they
offer an exact non-perturbative treatment \cite{Mussardo:1992uc,Dorey:1996gd}. 
Their exact bootstrap solution
does not start from any Lagrangian, rather it determines the scattering
matrices of the particles from such consistency requirements as unitarity
and crossing symmetry assuming maximal analyticity. In contrast to
the ultraviolet (UV) description based on the Lagrangian the infrared
(IR) formulation relies on the particle masses and the scattering matrices.
In the simplest case of the scaling Lee-Yang model there is only one
type of particle with a given mass and the scattering matrix is a
simple CDD factor without any parameter \cite{Cardy:1989fw}. The procedure to connect the
large scale IR scattering theory to a small scale UV Lagrangian formulation
is to put the system in a finite size and calculate an interpolating
quantity, such as the ground-state energy, exactly.

The Thermodynamic Bethe Ansatz (TBA) equation \cite{Zamolodchikov:1989cf}
describes the ground state
energy from the IR side by summing up all the vacuum polarization
effects. This is a nonlinear integral equation depending on the scattering
matrix and the masses of the particles. Unfortunately the TBA equation 
does not allow any systematic analytic small volume expansion. 
Nevertheless, the central charge of the UV limiting theory and the bulk energy constant
can be extracted exactly. The central charge basically identifies
the UV conformal field theory (CFT), which is perturbed with 
relevant operators.
Demanding the integrability of the perturbation leaves a
few choices, from which the one matching with the IR description can be
easily singled out. The identification between the UV perturbed CFT (pCFT)
Lagrangian and the IR scattering theory boils down to the relation
between the mass of the fundamental particle and the strength of the
perturbation. This relation is called the mass-coupling relation and 
is a real challenge to calculate in any integrable model. This 
relation is of fundamental significance as it also gives the vacuum expectation
values of the perturbing operators, which  contain
all the non-perturbative information which is not captured by the pCFT
\cite{Zamolodchikov:1990bk,Lukyanov:1996jj}. 

In order to calculate the mass-coupling relation one typically  embeds 
the theory into a larger model with extra symmetries.
After introducing some type of magnetic field coupled to the extra
conserved current the TBA equations can be linearized and expanded
systematically. Comparing the result with the analogous 
perturbative expansion on the Lagrangian side the relation between the masses 
and the parameters of the Lagrangian can be established.
This route was followed for the O(3) \cite{Hasenfratz:1990zz} and 
sine-Gordon models \cite{Zamolodchikov:1995xk} 
and has been extended for many other integrable models 
\cite{Hasenfratz:1990ab,Forgacs:1991rs,Forgacs:1991nk,
Balog:1992cm,Fateev:1992tk,Fateev:1993av,Hollowood:1994np,Evans:1994sy,Evans:1994sv}.
(For a different route, see \cite{Fateev:1991bv}.)
None of these models, however, contains integrable
perturbations with more 
 than one mass scale. Even though the models have multi-parameters and/or
a non-trivial spectrum, the mass ratios are encoded in the S-matrix.

Such integrable models with multiple  mass scales are obtained by
more general cosets with rank higher than the $su(2)$ cosets 
of  minimal models.  The homogeneous sine-Gordon (HSG) models, 
which are perturbed generalized parafermionic CFTs,
provide a simple class \cite{FernandezPousa:1996hi,FernandezPousa:1997zb,
FernandezPousa:1997iu,Miramontes:1999hx,CastroAlvaredo:1999em,Dorey:2004qc}. 
They are also distinct in that they are generically 
parity asymmetric, possess unstable particles and exhibit cross-over phenomena 
due to the multi-scales \cite{CastroAlvaredo:2000ag,CastroAlvaredo:2000nr,Dorey:2004qc}. 

Moreover, the free energy of the HSG models gives the strong-coupling 
gluon scattering
amplitudes of the four-dimensional maximally supersymmetric Yang-Mills theory 
(${\cal N}=4$ SYM) through their TBA equations  
\cite{Alday:2007hr,Alday:2009yn,Alday:2009dv,Alday:2010vh,Hatsuda:2010cc}.
Based on this fact, an analytic expansion of the amplitudes has been investigated 
around a certain kinematic point corresponding to the UV limit of the HSG models
via bulk and boundary pCFT  for the free energy and for the Y-functions
\cite{Hatsuda:2010vr,Hatsuda:2011ke,Hatsuda:2011jn,Hatsuda:2012pb,Hatsuda:2014vra}.
In order to make this expansion powerful an explicit connection is needed 
between the expressions given in terms of the IR/TBA data and 
those obtained analytically in terms of the UV/pCFT data. This missing link
would be provided by the mass-coupling relation. 

In this paper, we thus initiate a systematic study of the mass-coupling relation 
of multi-scale integrable models. Our main focus is on the simplest among such 
models, which is the perturbed  $\frac{su(3)_2}{u(1)^2}$ theory.

The paper is organized as follows: In Section \ref{sec.HSGpCFT} we describe the homogenous sine-Gordon
models as perturbed CFTs. We start by recalling the perturbed coset representation of the
theory. We then exploit the fact that it has an alternative coset representation, which can be
equivalently rewritten in terms of the projected product of  minimal models. 
We use this minimal model representation to confirm 
the modular invariant partition function and to identify its
integrable perturbations. The latter 
is done by constructing spin 1 conserved charges and by showing
the existence of spin 3 charges. The pCFT description allows us 
to calculate order by order the
ground-state energy, which is an analytical small volume expansion. Section \ref{sec.scat} collects the
analogous information about the model for large volumes. The model is defined by its particle
content and their scattering matrices. These data can be used to derive TBA integral equations
for the ground-state energy valid at any finite size. The operators are defined by their form factors.
We identify the IR basis of the perturbing fields and the densities of conserved 
spin 1 charges. We then use in Section \ref{sec.AnMCrel} form
factor perturbation theory to relate the IR basis 
to the UV basis by the mass-coupling
relation. Finally, using a generalization of the $\Theta$ sum rule Ward identities coming from
the conservation laws, we derive differential equations for the mass-coupling relations,
which we solve explicitly in terms of hypergeometric functions. These analytical mass-coupling
relations are compared in Section \ref{sec.NumMCrel} to the ones which we obtain 
by numerically solving the TBA
equations. As we find complete agreement we use the mass-coupling relation in Section \ref{sec.VEV} to
analyze the vacuum expectation values of the perturbing fields and conclude in Section \ref{sec.concl}.
To make the relatively long paper readable the technical details are relegated to various
appendices.
Our conventions are summarized in Appendix \ref{app.convention}.
The exact mass-coupling relation presented in Section \ref{subsec.solDE}
has been announced in \cite{Bajnok:2015eng}.

\section{Homogeneous sine-Gordon model as a perturbed CFT}
\label{sec.HSGpCFT}

In this section we describe the
simplest HSG model with multi-coupling deformations, 
namely the $su(3)_2/u(1)^2$ HSG model,
as a perturbed CFT. We start by introducing the model as integrable perturbations of the coset 
$su(3)_2/u(1)^2$ CFT. We then discuss in some detail the representation 
of the same model in terms of the projected product of minimal models. 
This second representation is useful as 
the structure constants and the correlation functions of minimal models are 
all well-known. We construct the conserved currents, 
and analyze the ground state energy from the pCFT point of view.
The results on conserved currents and the symmetries of  
the ground state energy will be important later in the discussion 
of the exact mass-coupling relation in Section \ref{sec.AnMCrel}.

\subsection{Coset representation}
\label{subsec.coset}

The homogeneous sine-Gordon models 
\cite{FernandezPousa:1996hi,FernandezPousa:1997zb,FernandezPousa:1997iu,
Miramontes:1999hx,CastroAlvaredo:1999em,Dorey:2004qc}
are obtained by integrable deformations 
of the $g_{k}/u(1)^{r_{g}}$ coset   CFTs \cite{Fateev:1985mm,Gepner:1986hr}, 
where $k$ is the level, $g$
is a simple compact Lie algebra and $r_{g}$ is its rank. The deforming
term consists of the weight-$0$ primary fields in the adjoint representation
of $g$, which are $r_g$ degenerate in the holomorphic sector.
Combining them with the antiholomorphic sector, 
the complete basis can be denoted as $\Phi_{ij}$ $(i,j=1,\dots,r_{g})$.
The actions of the HSG models take the form 
\begin{equation}
S_{\rm HSG}= S_{\rm CFT} - \int d^{2}x\, \mc{L}_{\rm pert}
    \comma\quad \mc{L}_{\rm pert} = 
    \sum_{i,j=1}^{r_{g}} \nu_{ij} \Phi_{ij}\comma \label{eq.HSGaction}
\end{equation}
where $S_{\rm CFT}$ is the action of the coset CFT or the gauged  
Wess--Zumino--Novikov--Witten model.
The left/right conformal dimensions of the deforming fields $\Phi_{ij}$ are all the same. 
Denoting them by $(h,h)$, those of the couplings $\nu_{ij}$ are $(1-h,1-h)$.
The couplings are factorized as
\beq
  \label{eq.mufact}
   \nu_{ij} = \lambda_{i}\bar{\lambda}_{j} \period
\eeq
These dimensionful coupling constants are not renormalized 
in the perturbative CFT scheme and hence are physical themselves \cite{Zamolodchikov:1990bk, CF}.
Due to the invariance under a rescaling 
$(\lambda_{i},\bar{\lambda}_{j})\to(\alpha\lambda_{i},\alpha^{-1}\bar{\lambda}_{j})$,
the number of the independent couplings $(\lambda_{i},\bar{\lambda}_{j})$
is $2r_{g}-1$. Thus, for $r_{g}>1$, the HSG models are distinct
in that they remain integrable under multi-coupling deformations.

In the UV regime, one can investigate the HSG models
by regarding them as perturbed CFTs. A useful fact in this respect
is that coset CFTs often have equivalent representations by other
cosets. In the case of $g=su(n)$, which is relevant to our discussion,
one has \cite{Bagger:1988px,Bouwknegt:1992wg} 
\begin{eqnarray}
\frac{su(n)_{k}}{u(1)^{n-1}} & \cong & \frac{su(k)_{1}^{(1)}\times su(k)_{1}^{(2)}
\times\dots\times su(k)_{1}^{(n)}}{su(k)_{n}}\nn\label{coset}\\
& \cong & \frac{su(k)_{1}\times su(k)_{1}}{su(k)_{2}}\times\frac{su(k)_{2}
\times su(k)_{1}}{su(k)_{3}}\times\dots\times\frac{su(k)_{n-1}\times su(k)_{1}}{su(k)_{n}}\comma
\end{eqnarray}
up to identifications of the common factors in the denominators and
the numerators. The superscripts in $su(k)_{1}^{(p)}$ just express 
that it is the $p$-th factor. Since the unitary minimal model with
the central charge $c_{m}=1-6/m(m+1)$ is represented by the $su(2)$
diagonal coset as $\calM_{m+2,m+3}=su(2)_{m}\times su(2)_{1}/su(2)_{m+1}$,
the second line in (\ref{coset}) implies for $k=2$ that 
\begin{eqnarray}
\frac{su(n)_{2}}{u(1)^{n-1}} & = & \mathbb{P} \left( \calM_{3,4}\times\calM_{4,5}
\times\dots\times\calM_{n+1,n+2} \right )\period\label{cosetMM}
\end{eqnarray}
We have explicitly indicated by $\mathbb{P}$ that the product is the projected one
due to the  identifications implicit in (\ref{coset}).

In the rest of the present paper we study the case of $n=3$, which
corresponds to the simplest HSG model possessing all the characteristic
features mentioned above. The $su(3)_{2}/u(1)^{2}$ coset CFT in this
case has nine chiral primary fields.
Their conformal dimensions are given by $h=0$ (identity), $1/10,1/2,3/5$
with the multiplicities $1,3,3,2$, respectively. The fields of dimension $3/5$
form the perturbing fields $\Phi_{ij}$.
In each set of three
fields with $h=1/10$ or $1/2$, they are related to each other by
the $\ZZ_{3}$ symmetry of $su(3)$. At level $k=2$, only the diagonal
modular invariant may be allowed, which is expressed by the string
functions of $su(3)_{2}$ (see Appendix \ref{app.b}). Properties of the $su(3)_{2}/u(1)^{2}$
coset theory have been summarized in \cite{Ardonne:2006fk}.

According to (\ref{cosetMM}), the $su(3)_{2}/u(1)^{2}$ coset CFT
is represented equivalently by a projected product of the Ising ($\calM_{3,4}$) and
the tricritical Ising ($\calM_{4,5}$) CFT, and has central charge
$c=\frac{6}{5}=\frac{1}{2}+\frac{7}{10}$, the sum of those of $\calM_{3,4}$ and $\calM_{4,5}$.
In each of the chiral sectors,
the spectrum of $\calM_{3,4}$ consists of the fields with $h=0$,
$1/16$ and $1/2$, respectively, whereas that of $\calM_{4,5}$ consists
of the fields with $h=0,3/80,1/10,7/16,3/5$ and $3/2$. All the multiplicities
are 1. The identification of the $su(2)_{2}$ factor implies that
only certain combinations of the fields in $\calM_{3,4}\times\calM_{4,5}$
appear in the spectrum of $su(3)_{2}/u(1)^{2}$. The possible combinations
are identified by the character decomposition of the coset CFT in
terms of the Virasoro characters and the affine $su(2)$ characters. Denoting
the primaries with conformal dimension $h$ by $\ket{h}$, the result
including the multiplicities reads 
\begin{equation}
{\textstyle \bigl(\vert0\rangle+\vert\frac{1}{2}\rangle\bigr)_{\calM_{3,4}}
\times\bigl(\vert0\rangle+\vert\frac{1}{10}\rangle+\vert\frac{3}{5}\rangle
+\vert\frac{3}{2}\rangle\bigr)_{\calM_{4,5}}
+\ 2\times\bigl(\vert\frac{1}{16}\rangle\bigr)_{\calM_{3,4}}
\times\bigl(\vert\frac{7}{16}\rangle+\vert\frac{3}{80}\rangle\bigr)_{\calM_{4,5}}\period}
\label{FieldContent}
\end{equation}
Up to the states which can be interpreted as descendants in terms
of the larger $su(3)_{2}/u(1)^{2}$ algebra, the above chiral spectrum
indeed agrees with that of the $su(3)_{2}/u(1)^{2}$ coset theory.

Moreover, the modular invariant of the $su(3)_{2}/u(1)^{2}$ theory
is expressed by the Virasoro characters of $\calM_{3,4}$ and $\calM_{4,5}$
and it has to be compatible with the field content (\ref{FieldContent}).
As shown shortly, one can construct this modular invariant by starting
directly from the Virasoro characters. For definiteness, we summarize
the relations among the $su(2)_{k}$ and the Virasoro characters,
and the $su(3)_{2}$ string functions in Appendix \ref{app.b}.

\subsection{Minimal models product representation}
\label{subsec.MMrep}

In this subsection we consider the $su(3)_{2}/u(1)^2$ homogeneous sine-Gordon
model as perturbations of the projected product of minimal models
$ \mathbb{P}(\mathcal{M}_{3,4}\times\mathcal{M}_{4,5})$.
We first give a description of the chiral algebras and build up the
Hilbert space from their highest weight representations by choosing
the relevant modular invariant partition function. 
In this picture we easily identify
a multi-parameter family of integrable perturbations by demanding the
existence of higher spin conserved charges. In particular, integrability
ensures that the perturbing operators themselves 
are components of conserved currents. 
Note that the conserved charges correspond to off-critical 
deformations of some of the elements of the enveloping algebra 
of the chiral algebra.

\subsubsection{The chiral algebra}

The chiral algebra of $ \mathcal{M}_{3,4}\times\mathcal{M}_{4,5}$
contains two commuting Virasoro algebras 
\begin{equation}
[L_{n}^{(i)},L_{m}^{(i)}]=(n-m)L_{n+m}^{(i)}+\frac{c^{(i)}}{12}(n^{3}-n)\delta_{n+m}\, , \qquad 
[L_{n}^{(1)},L_{m}^{(2)}]=0 \, ,
\end{equation}
with central charges $c^{(1)}=\frac{1}{2}$ and $c^{(2)}=\frac{7}{10}$,
such that the total Virasoro generator is 
\begin{equation}
L_{n}=L_{n}^{(1)}+L_{n}^{(2)}\ .
\end{equation}

\paragraph*{The Ising part}

As the Ising model is the free massless fermion theory, we may introduce
the fermion field $\psi(z)=\sum_{n}z^{-n-1/2}\psi_{n}$, 
where $n\in\ZZ+\frac{1}{2}$ for the Neveu-Schwarz (NS) sector and $n\in\ZZ$ 
for the Ramond (R) sector.
The modes $ \psi_n $ have anticommutation relations 
\begin{equation}
\{\psi_{n},\psi_{m}\}=\delta_{n+m} \, ,
\end{equation}
such that 
\begin{equation}
L^{(1)}(z)=\frac{1}{2}:\partial\psi(z)\psi(z):\ ,
\end{equation}
where $:\ :$ denotes normal ordering. Let $\ket{0}$ denote the vacuum vector, satisfying 
$\psi_n\ket{0}=0$ ($n>0$).
The two representations corresponding to the highest weight vectors
$\ket{0}$ and $\ket{\frac{1}{2}}=\psi_{-\frac{1}{2}}\ket{0}$ form
the vacuum representation of the free fermion algebra. This is the
NS representation with half-integer moding. The highest
weight representation built on $\ket{\frac{1}{16}}$ is the
R representation with integer moding.

\paragraph*{The tricritical Ising part}

The presence of the field with conformal dimension $h=3/2$ in the
tricritical Ising model indicates that it is actually a superconformal model with 
\begin{equation} 
[L_{n}^{(2)},G_{m}]=\Bigl(\frac{n}{2}-m\Bigr)G_{n+m} \, , 
\end{equation}
\begin{equation} 
\{G_{n},G_{m}\}=2L_{n+m}^{(2)}+\frac{c^{(2)}}{3}\biggl(n^{2}-\frac{1}{4}\biggr)\delta_{n+m}\ .
\end{equation}
The two Virasoro modules built over $\ket{0}$ and $\ket{\frac{3}{2}}=G_{-\frac{3}{2}}\ket{0}$
form the vacuum module of the superconformal algebra, while the one
built over $\ket{\frac{1}{10}}$ and $\ket{\frac{3}{5}}=G_{-\frac{1}{2}}\ket{\frac{1}{10}}$
the NS type highest weight representation. The R representations of
the superconformal algebra are built on $\ket{\frac{7}{16}}$ and
$\ket{\frac{3}{80}}$.

The chiral algebra of the product picture is generated by the fields
$\mathcal{A}=\{\psi,L^{(2)},G\}$. 
In particular, it contains three spin 2 chiral fields: $L^{(1)}(z)$,
$L^{(2)}(z)$, and $L^{(3)}(z)=\psi(z) G(z)$, which will play a central
role in our considerations.
Below, we search for the relevant modular invariant partition
function on the torus in this picture which accommodates 4 fields with dimensions 
$(\frac{3}{5},\frac{3}{5})$
required by the homogeneous sine-Gordon models.

\subsubsection{The Hilbert space of the product model}

To construct the modular invariant, we start from the vacuum module of $\mathcal{A}$.
\begin{equation}
(\chi_{00}+\chi_{\frac{1}{2}0}+\chi_{0\frac{3}{2}}+\chi_{\frac{1}{2}\frac{3}{2}})(\bar{\chi}_{00}
+\bar{\chi}_{\frac{1}{2}0}+\bar{\chi}_{0\frac{3}{2}}+\bar{\chi}_{\frac{1}{2}\frac{3}{2}}) \, ,
\end{equation}
where 
$\chi_{hh'}=\chi_{h}^{(1)}\chi_{h'}^{(2)}$ 
denotes the Virasoro
character of the product representation. The NS representation  of
the chiral algebra is given by
\begin{equation}
(\chi_{0\frac{1}{10}}+\chi_{0\frac{3}{5}}+\chi_{\frac{1}{2}\frac{1}{10}}
+\chi_{\frac{1}{2}\frac{3}{5}})(\bar{\chi}_{0\frac{1}{10}}+\bar{\chi}_{0\frac{3}{5}}
+\bar{\chi}_{\frac{1}{2}\frac{1}{10}}+\bar{\chi}_{\frac{1}{2}\frac{3}{5}}) \, ,
\end{equation}
which contains 4 fields with dimensions 
$(\frac{3}{5},\frac{3}{5})$
as required by the coset correspondence. The sum of the characters
of these representation spaces is almost modular invariant. 
It is invariant with respect to the modular $S$ transformation, but
fermionic elements, where the difference of the left and the right
dimensions is half integer, pick up a sign for $T$ transformation.
The result of this action can be concisely written as 
\begin{equation}
(\chi_{00}-\chi_{\frac{1}{2}0}-\chi_{0\frac{3}{2}}+\chi_{\frac{1}{2}\frac{3}{2}})
(\bar{\chi}_{00}-\bar{\chi}_{\frac{1}{2}0}-\bar{\chi}_{0\frac{3}{2}}
+\bar{\chi}_{\frac{1}{2}\frac{3}{2}}) \, , 
\end{equation}
on the vacuum sector and as 
\begin{equation}
(\chi_{0\frac{1}{10}}-\chi_{0\frac{3}{5}}-\chi_{\frac{1}{2}\frac{1}{10}}
+\chi_{\frac{1}{2}\frac{3}{5}})(\bar{\chi}_{0\frac{1}{10}}
-\bar{\chi}_{0\frac{3}{5}}-\bar{\chi}_{\frac{1}{2}\frac{1}{10}}+\bar{\chi}_{\frac{1}{2}\frac{3}{5}}) \, , 
\end{equation}
on the NS sector. The modular $S$ transformation acting on these
later characters produces the characters of the twisted R sector as
\begin{equation}
4\chi_{\frac{1}{16}\frac{3}{80}}\bar{\chi}_{\frac{1}{16}\frac{3}{80}}
+4\chi_{\frac{1}{16}\frac{7}{16}}\bar{\chi}_{\frac{1}{16}\frac{7}{16}} \, .
\end{equation}
The full modular invariant partition function can be obtained by summing up
them all. Actually since every space appears twice, we take its half and get the
following modular invariant partition function: 
\begin{eqnarray}
\label{eq.cosetchara}
Z & = & 2\chi_{\frac{1}{16}\frac{3}{80}}\bar{\chi}_{\frac{1}{16}\frac{3}{80}}
+2\chi_{\frac{1}{16}\frac{7}{16}}\bar{\chi}_{\frac{1}{16}\frac{7}{16}}\nn\\
 &  & +(\chi_{00}+\chi_{\frac{1}{2}\frac{3}{2}})(\bar{\chi}_{00}
 +\bar{\chi}_{\frac{1}{2}\frac{3}{2}})+(\chi_{\frac{1}{2}0}
 +\chi_{0\frac{3}{2}})(\bar{\chi}_{\frac{1}{2}0}+\bar{\chi}_{0\frac{3}{2}})\nn\\
 &  & +(\chi_{0\frac{1}{10}}+\chi_{\frac{1}{2}\frac{3}{5}})
 (\bar{\chi}_{0\frac{1}{10}}+\bar{\chi}_{\frac{1}{2}\frac{3}{5}})
 +(\chi_{0\frac{3}{5}}+\chi_{\frac{1}{2}\frac{1}{10}})
 (\bar{\chi}_{0\frac{3}{5}}+\bar{\chi}_{\frac{1}{2}\frac{1}{10}}) \, .
\end{eqnarray}
The chiral algebra
of the coset conformal field theory is larger than that of the product of
minimal models, thus the diagonal modular invariant partition function
on the coset side is not diagonal in terms of the Virasoro characters. 
Rather, it contains off-diagonal terms signaling the presence of the larger
chiral current algebra. 

From this expression one can easily read off the field content of
the model, which additionally to the vacuum sector contains 3 fields
with highest weights
$\left(\frac{1}{10},\frac{1}{10}\right)$, 3 fields with
highest weights $\left(\frac{1}{2},\frac{1}{2}\right)$ and 4 fields
with highest weights $\left(\frac{3}{5},\frac{3}{5}\right)$ required from the coset point
of view. This is the model we would like to perturb with the $\left(\frac{3}{5},\frac{3}{5}\right)$
fields, whose corresponding vectors we denote by
\begin{equation}
{\textstyle \vert\Phi_{ij}\rangle=\psi_{-\frac{1}{2}}^{(i)}\bar{\psi}_{-\frac{1}{2}}^{(j)}\vert\Phi\rangle
  \quad  (i,j=1,2) } \, ,
\end{equation}
\begin{equation}
\textstyle
\vert\Phi\rangle=\vert\frac{1}{10},\frac{1}{10}\rangle \, , \quad \quad 
\psi^{(1)}=\psi \, ,  \quad 
\quad\psi^{(2)}=\sqrt{5}G \, ,
\end{equation}
such that they form an orthonormal basis
\begin{equation}
\langle\Phi_{ij}\vert\Phi_{kl}\rangle=\delta_{ik}\delta_{jl} \, .
\end{equation}
Finally, we note that the actual 
chiral algebra is the remnant of the current algebra 
of the coset theory, which is larger than the one 
generated by  $\psi(z),L^{(2)}(z),G(z)$. The missing fields are 
related to the other two fermions  with $h=\frac{1}{2}$ which
appear in $2\chi_{\frac{1}{16}\frac{7}{16}}
\bar{\chi}_{\frac{1}{16}\frac{7}{16}}$ as shown in 
Section \ref{subsec.coset} and Appendix \ref{app.b}.

\subsubsection{Perturbation and conserved charges}
\label{cons.charge}

Given the Hilbert space of the model in the product picture,
let us move on to a discussion on the conserved charges.
As we are working with a smaller chiral algebra 
we do not expect to find all conserved charges in this picture. 
We start from the perturbed action of the form (\ref{eq.HSGaction}), 
not assuming (\ref{eq.mufact}).
In the present case $g_k$ = $su(3)_2$, $r_g=2$ and 
the left/right dimensions of $\Phi_{ij}$ and $\nu_{ij}$ are
$(h,h)=\left(\frac{3}{5},\frac{3}{5}\right)$ and $(1-h,1-h)$, respectively.
It turns out below that the couplings $\nu_{ij}$ must factorize as in (\ref{eq.mufact})
to ensure the integrability of the model. 

Integrability requires an infinite
number of conserved charges. 
In the conformal field theory, 
where all couplings vanish, $\nu_{ij}=0$, any differential normal-ordered polynomial
of the generating fields of the chiral algebra corresponds to a conserved
charge. Indeed, taking a representative $\Lambda(z)$, it depends
only on $z$ and $\bar{\partial}\Lambda(z)=0$. After we introduce
the perturbation this is no longer true, but we can systematically
calculate the corrections 
\begin{equation}
\bar{\partial}\Lambda(z,\bar{z})=
\nu_{ij}\Theta_{ij}(z,\bar{z})+\nu_{ij}\nu_{kl}\Theta_{ijkl}(z,\bar{z})
+\dots \, . \label{eq:CPT}
\end{equation}
What is nice about the perturbed rational unitary  conformal field theories is that,
due to the discrete and nonnegative set of the allowed scaling weights, the conformal
perturbation theory terminates with a finite number of terms only.
This can be seen by comparing the dimensions of the two sides of eq.
(\ref{eq:CPT}). Let us assume that the conformal dimension of $\Lambda$ is
$(s,0)$ with $s$ being a positive integer. Associating the dimension
$(\Delta,\bar{\Delta})$ to $\Theta_{ij}$ the comparison gives $(s,1)=(1-h+\Delta,1-h+\bar{\Delta})$
at first order, which means that $\Delta=h+s-1$ and $\bar{\Delta}=h$,
i.e. $\Theta_{ij}$ is a level $s-1$ left descendant of the $\Phi_{ij}$s.
Inspecting the second order perturbation we find that $\bar{\Delta}=2h-1=1/5$.
But there are no fields with this dimension, so the second order perturbation
vanishes. As there are no fields with negative dimensions either,
all higher order terms vanish and we conclude that the first order
perturbation is actually {\em exact}.

Clearly we cannot introduce a total derivative for $\Lambda$ as its integral vanishes
and does not give rise to any conserved charge. Thus 
the existence of an off-critical conserved current requires that $\Lambda$ is not, but the level
$s-1$ descendant is a total derivative: 
\begin{equation}
\bar{\partial}\Lambda(z,\bar{z})=\nu_{ij}\Theta_{ij}(z,\bar{z})=\nu_{ij}\partial A_{ij}(z,\bar{z})
\, .
\end{equation}
If we are interested only in the existence of the conserved charge,
and not its explicit form, we only have to compare the dimension of the nonderivative 
operators at level $s$  in the chiral 
algebra to the dimension of the level $s-1$ derivative descendants of $\Phi_{ij}$. If the 
former is larger, then we can construct a conserved charge.
The argument based on this is called the counting
argument \cite{Zamolodchikov:1987zf,Zamolodchikov:1989zs}. It is presented in Appendix \ref{app.count}.

One can actually do a better job and determine explicitly the  linear combinations
\begin{equation}
\Lambda=\alpha_{1}L^{(1)}+\alpha_{2}L^{(2)}+\alpha_{3}L^{(3)}\ ,
\end{equation}
with some constants $\alpha_1, \alpha_2$ and $\alpha_3$,
which remain conserved under the perturbation. Doing the first order perturbative
calculation (see the master formula in Appendix \ref{subsect.master})
\begin{equation}
\bar{\partial}\Lambda(z,\bar{z})
=-\pi\nu_{ij}\oint_{z}\frac{dw}{2\pi \ii}
\left(\Lambda(z)\Phi_{ij}(w,\bar{z})\right) \, ,
\end{equation}
we need the OPE 
\begin{equation}
\Lambda(z)\Phi_{ij}(w,\bar{w})
=\frac{(\Lambda\Phi_{ij})_{-2}(w,\bar{w})}{(z-w)^{2}}
+\frac{(\Lambda\Phi_{ij})_{-1}(w,\bar{w})}{z-w}+\dots \, ,
\end{equation}
to obtain
\begin{equation}
\bar{\partial}\Lambda(z,\bar{z})=-\pi\nu_{ij}
\left[\partial((\Lambda\Phi_{ij})_{-2}(z,\bar{z}))-(\Lambda\Phi_{ij})_{-1}(z,\bar{z})\right] \, .
\end{equation} 

Writing formally $\Phi_{ij}=\phi_i \bar{\phi}_j$ the OPEs with the (chiral part of the) perturbing fields are 
calculated to be
\begin{equation}
\begin{split}
L^{(1)}(z)\phi_1(w)&=
\frac{1}{2}\frac{\phi_1(w)}{(z-w)^2}+\frac{\partial\phi_1(w)}{z-w}
-\frac{\phi_x(w)}{z-w}+{\rm O}(1)\\
L^{(1)}(z)\phi_2(w)&={\rm O}(1)\\
L^{(2)}(z)\phi_1(w)&=
\frac{1}{10}\frac{\phi_1(w)}{(z-w)^2}
+\frac{\phi_x(w)}{z-w}+{\rm O}(1)\\
L^{(2)}(z)\phi_2(w)&=
\frac{3}{5}\frac{\phi_2(w)}{(z-w)^2}+\frac{\partial\phi_2(w)}{z-w}
+{\rm O}(1)\\
L^{(3)}(z)\phi_1(w)&=
\frac{1}{\sqrt{5}}\frac{\phi_2(w)}{(z-w)^2}
+\frac{\sqrt{5}}{3}\frac{\partial\phi_2(w)}{z-w}
+{\rm O}(1)\\
L^{(3)}(z)\phi_2(w)&=
\frac{1}{\sqrt{5}}\frac{\phi_1(w)}{(z-w)^2}
+\frac{1}{\sqrt{5}}\frac{\partial\phi_1(w)}{z-w}
+\frac{4}{\sqrt{5}}\frac{\phi_x(w)}{z-w}+{\rm O}(1).
\end{split}
\end{equation} 
Here $\phi_x$ is a non-derivative field.  

Clearly the total energy and momentum
is always conserved:
\begin{equation}
\label{eq.EMconsv}
\bar{\partial}L(z,\bar{z})=\bar{\partial}(L^{(1)}(z,\bar{z})+L^{(2)}(z,\bar{z}))
=\frac{2\pi}{5}\partial\left(\nu_{ij}\Phi_{ij}(z,\bar{z})\right) \, \period
\end{equation}
Combining $L^{(1)}$ and $L^{(3)}$ we demand the vanishing of the
non-derivative term, which leads to 
\begin{equation}
\alpha_{1}=\frac{4\alpha_{3}}{\sqrt{5}}\frac{\nu_{21}}{\nu_{11}}
=\frac{4\alpha_{3}}{\sqrt{5}}\frac{\nu_{22}}{\nu_{12}} \, .
\end{equation}
The compatibility of these two equations implies 
$\nu_{12}\nu_{21}=\nu_{11}\nu_{22}$, which is equivalent to the 
factorization 
of the coefficients of the perturbation  
as in (\ref{eq.mufact}).
Actually from this factorization it follows that we can search for the
conserved charges separately at each chiral half as the other chiral
half behaves merely as a spectator. 

The conservation law now takes the form 
\begin{equation}
\label{eq.Consv2}
\bar{\partial}\left(\alpha_{1}L^{(1)}(z,\bar{z})+\alpha_{3}L^{(3)}(z,\bar{z})\right)
=\partial\left(v_{1}\Psi_{1}(z,\bar{z})+v_{2}\Psi_{2}(z,\bar{z})\right) \, , 
\quad\Psi_{i}=\bar{\lambda}_{j}\Phi_{ij}
\, , 
\end{equation}
with a possible normalization 
\begin{equation}
\alpha_{1}=2 \,  , \quad 
\alpha_{3}=\frac{\sqrt{5}}{2} \frac{\lambda_1}{\lambda_2}\, , \quad 
v_{1}=\pi\lambda_{1} \,  , \quad 
v_{2}=\frac{\pi}{3}\frac{\lambda_{1}^2}{\lambda_2} \, .
\end{equation}
By the left/right symmetry of the problem we also have a conservation
law by replacing each quantity with its bar version: 
\begin{equation}
\partial\left(\bar{\alpha}_{1}\bar{L}^{(1)}(z,\bar{z})+\bar{\alpha}_{3}
\bar{L}^{(3)}(z,\bar{z})\right)=\bar{\partial}\left(\bar{v}_{1}\bar{\Psi}_{1}(z,\bar{z})
+\bar{v}_{2}\bar{\Psi}_{2}(z,\bar{z})\right)
\, ,
\end{equation}
where $\bar{\alpha}_1=2$, $\bar{\alpha}_3=\sqrt{5}\bar{\lambda}_1/(2\bar{\lambda}_2)$ 
and $\bar{v}_{1}=\pi\bar{\lambda}_{1}$, $ \bar{v}_{2}=\pi \bar{\lambda}_{1}^2/(3
\bar{\lambda}_2)$. In the following we do not write out explicitly the formulas which can 
be obtained by the left/right replacements. 

The results above on the conserved currents and corresponding charges 
turn out to be important later.
 For future extension to the $su(n)_{2}/u(1)^{n-1}$ theory,
the projected product of the minimal models
(\ref{cosetMM}) is further discussed in Appendix \ref{app.proj}.

\subsection{Ground state energy from perturbed CFT}
\label{subsec.cpt}

From the pCFT formulation of the model, we can derive 
 pieces of information on the ground state energy,
which are used in the later analyses.
We consider  the dimensionless
ground state energy $F(L)=\frac{L}{2\pi}E_{0}(L)$, which can be expanded
at small cylinder circumference $L$ as 
\begin{equation}
F(L)=-\frac{c}{12}+\sum_{n=1}^{\infty}F_{n}L^{n(2-2h)}\ ,\label{eq.uvexp}
\end{equation}
where $c=\frac{6}{5}$ is the central charge of the $su(3)_{2}/u(1)^{2}$
coset CFT and $h=\frac{3}{5}$ is the dimension of $\Phi_{ij}$.
The perturbative coefficients are 
\begin{equation}
F_{n}=
\frac{-1}{n!}(2\pi)^{1+2n(h-1)}
\int\langle0\vert \prod_{k=1}^n \lambda_{i_{k}}\bar{\lambda}_{j_{k}}\Phi_{i_{k}j_{k}}(z_{k},\bar{z}_{k})
 \vert0\rangle_c 
\prod_{k=2}^{n}(z_{k}\bar{z}_{k})^{(h-1)}{\rm d}^2 z_{k} \, ,
\end{equation}
where the subscript in $\langle\cdot \rangle_c$  
stands for the connected part and $z_{1}=\bar{z}_{1}=1$. 

The operators $\Phi_{ij}$ and the identity $\id$ form a closed set under operator
product expansion (OPE). 
Formally, we can choose a basis $\phi_i(z), \bar\phi_j(\bar{z})$
of the  fields of dimension $(\frac{3}{5},0)$ and $(0,\frac{3}{5})$ and may write
\beq
  \Phi_{ij}(z,\bar{z})\equiv\phi_{i}(z)\bar{\phi}_{j}(\bar{z}) \, ,
\eeq
such that the OPE rules read
\cite{Ardonne:2006fk} 
\begin{eqnarray}
\phi_{1}\phi_{1}=\id-\sqrt{2C}\phi_{2}\ , & \quad\phi_{1}\phi_{2}=-\sqrt{2C}\phi_{1}\ , & \quad\phi_{2}\phi_{2}=\id+\sqrt{2C}\phi_{2}\ ,\label{eq.ope}
\end{eqnarray}
where 
\begin{equation}
C=\frac{1}{3}\gamma^{\frac{1}{2}}\left(\frac{4}{5}\right)\gamma^{\frac{3}{2}}\left(\frac{2}{5}\right)\ ,
\end{equation}
and $\gamma(x)=\Gamma(x)/\Gamma(1-x)$.
In the above formulas, only the leading terms are shown, and 
the dependence on spacetime variables is suppressed. 
These OPEs are invariant under
rotations of $\phi_{i}$ by $2\pi/3$, which form a $\ZZ_{3}$ group,
and under the reflection $\phi_{1}\to-\phi_{1}$, $\phi_{2}\to\phi_{2}$.
These transformations generate the symmetric group $S_{3}$, which
corresponds to the Weyl reflection group of $su(3)$. Due to this
symmetry of the model, $F_{n}(\lambda_{1},\lambda_{2},\bar{\lambda}_{1},\bar{\lambda}_{2})$
have to be invariant under 
the $S_3$ Weyl symmetry group generated by 
\begin{enumerate}
\item 
$\ZZ_3$ rotations: $\lambda_i  \to \omega_{ij} \lambda_j $ where $\omega_{ij}$ stands for 
   the $2\pi/3$ rotation  , 
\item
   reflection: $\lambda_{1}\to-\lambda_{1}$, $\lambda_{2}\to\lambda_{2}$  .
\end{enumerate}
The same applies separately to the variables $\bar{\lambda}_{i}$.
It is useful, therefore, to introduce the invariant polynomials 
\begin{equation}
p_{2}=\lambda_{1}^{2}+\lambda_{2}^{2},\qquad p_{3}=\lambda_{2}^{3}-3\lambda_{2}\lambda_{1}^{2}.\label{eq.p2}
\end{equation}
$p_{2}$ and $p_{3}$ generate all $S_{3}$-invariant polynomials
of $\lambda_{i}$, and the same applies, of course, to the similarly defined quantities
$\bar{p}_{2},\bar{p}_{3},\bar{\lambda}_{i}$.

In terms of these polynomials, we get for the perturbative
coefficients, 
\begin{equation}
F_{2}=C_{2}\, p_{2}\,\bar{p}_{2}\ ,\qquad\quad F_{3}=C_{3}\, p_{3}\,\bar{p}_{3}\ ,\label{eq.fw4}
\end{equation}
where $C_{2}$ and $C_{3}$ are constants which are read off from the integrals
of the two- and three-point functions of $\phi_{i}$ \cite{Zamolodchikov:1991vh} and
from the OPE coefficients in (\ref{eq.ope}): 
\begin{eqnarray}
C_{2} & = & -\frac{1}{4}(2\pi)^{\frac{2}{5}}\gamma\left(-\frac{1}{5}\right)\gamma^{2}\left(\frac{3}{5}\right)\\
C_{3} & = & -\frac{1}{24}\, C\,(2\pi)^{\frac{3}{5}}\gamma^{3}\left(\frac{3}{10}\right)\gamma^{-1}\left(\frac{9}{10}\right)\ .
\end{eqnarray}
$C_{2}$ is positive since $\gamma(-1/5)<0$.
If $\lambda_{i}$ are parametrized as 
$\lambda_{1}=\lambda\cos\varphi\ ,\lambda_{2}=\lambda\sin\varphi$,
then $p_{2}$ and $p_{3}$ take the form $p_{2}=\lambda^{2},p_{3}=-\lambda^{3}\sin3\varphi$,
and it can be seen immediately that 
\begin{equation}
0\le\frac{F_{3}^{2}}{F_{2}^{3}}\le\frac{C_{3}^{2}}{C_{2}^{3}}\ .\label{eq.ineq}
\end{equation}
In Section \ref{sec.NumMCrel}, the couplings $\lambda_i$ are determined 
through  (\ref{eq.p2}) and (\ref{eq.fw4})
by the numerical data of $F_2$ and $F_3$ which are obtained from the TBA equations.

From conformal perturbation theory it follows also 
that $F_{n}(\lambda_{1},\lambda_{2},\bar{\lambda}_{1},\bar{\lambda}_{2})$
are homogeneous polynomials of order $n$ both in $(\lambda_{1},\lambda_{2})$
and in $(\bar{\lambda}_{1},\bar{\lambda}_{2})$. Taking  into
consideration the $S_{3}$ symmetry described above, one finds that
$F_{4}$, $F_{5}$ and $F_{7}$ are determined up to constant factors
$C_{n}$, which are calculable, in principle, in pCFT: 
\begin{equation}
F_{n}=C_{n}\, p_{n}\,\bar{p}_{n} \qquad (n=4,5,7) \, , \label{eq.fwn}
\end{equation}
\begin{equation}
p_{4}=p_{2}^{2},\qquad p_{5}=p_{2}\, p_{3},\qquad p_{7}=p_{2}^{2}\, p_{3}\ .
\end{equation}
These relations imply that 
\begin{equation}
\frac{F_{4}}{F_{2}^{2}}=\frac{C_{4}}{C_{2}^{2}}\ ,\qquad\frac{F_{5}}{F_{2}F_{3}}=\frac{C_{5}}{C_{2}C_{3}}\ ,
\qquad\frac{F_{7}}{F_{2}^{2}F_{3}}=\frac{C_{7}}{C_{2}^{2}C_{3}}\ ,\label{eq.const}
\end{equation}
i.e.\ $\frac{F_{4}}{F_{2}^{2}}$, $\frac{F_{5}}{F_{2}F_{3}}$ and
$\frac{F_{7}}{F_{2}^{2}F_{3}}$ are constants.

For $F_{6}$ the discrete symmetries give the form 
\begin{equation}
F_{6}=C_{622}\, p_{2}^{3}\,\bar{p}_{2}^{3}+C_{633}\, p_{3}^{2}\,\bar{p}_{3}^{2}+C_{623}\, 
p_{2}^{3}\,\bar{p}_{3}^{2}+C_{632}\, p_{3}^{2}\,\bar{p}_{2}^{3}\ ,\label{eq.f6b}
\end{equation}
where $C_{622}$, $C_{633}$, $C_{623}$ and $C_{632}$ are constants,
and the symmetry between the holomorphic and antiholomorphic sectors
implies $C_{623}=C_{632}$.
 These relations are used for checking the precision of the numerical data from the TBA equations.

\section{Homogeneous sine-Gordon model as a scattering theory}
\label{sec.scat}

After the description of the model from the UV side,
we now turn to the description from the IR side. 
In the IR description, 
the physical masses (and the resonance parameter) are the fundamental 
variables of the system. To relate these to the perturbation couplings 
on the UV side is the main subject of this paper.
In the following, we first summarize the S-matrix and the TBA equation of the model. 
We then discuss the form factors. We find the form factors of the four dimension $3/5$
operators, which should be the IR counterpart of the perturbing operators $\Phi_{ij}$.
These results, together with those in the previous section, are used 
in the next section.

\subsection{S-matrix and TBA}

The spectrum of the HSG models contains stable solitonic particles
associated to the simple roots $\alpha_{a}$ of $g$. They are labeled
by two quantum numbers $(a,p)$ where $a=1,\dots,r_{g}$ and $p=1,\dots,k-1$,
and their masses are parametrized as $m_{a}^{(p)}=m_{a} \sin(\pi p/k) $.
The exact S-matrices describing the scattering
among those particles have been proposed in \cite{Miramontes:1999hx}.
These depend on further $r_{g}-1$ real parameters $\sigma_{ab}=-\sigma_{ba}$
assigned to each pair of neighboring nodes of the Dynkin diagrams.
$m_{a}$ and $\sigma_{ab}$ form a set of $2r_{g}-1$ independent
parameters, the number of which agrees with the one for the deformation
parameters. When the sum of the simple roots $\alpha_{a}+\alpha_{b}$
is a root, the S-matrix for the scattering of the corresponding particles
exhibits a pole where the rapidity variable $\theta$ coincides with
$\sigma_{ba}$. This is a resonance pole, signaling the formation
of an unstable particle associated with the root $\alpha_{a}+\alpha_{b}$.
Due to the resonance parameters $\sigma_{ab}$, the S-matrices are
not parity invariant. The existence of the resonance and the parity
non-invariance are characteristic of the HSG models. The scattering
properties feature their infrared (IR) behaviors.

For the $su(3)_2/u(1)^2$ HSG model with  $r_g = 2$,  $k=2$, there are 
 two self-conjugate particles of mass $m_1$ and $m_2$, which can take arbitrary values. 
We omit the superscript $p$ of $m_a^{(p)}$
as it can only be 1. There is only one resonance parameter
$\sigma_{12} =: \sigma$.
In this case, the two-particle S-matrix is given by \cite{Miramontes:1999hx}
\beq
\label{eq.s}
S_{12}(\theta - \sigma)=-S_{21}(\theta + \sigma)=
\tanh\left[\frac{1}{2}\left(\theta-\ii\frac{\pi}{2}\right)\right]
\, , \quad 
S_{11}(\theta)=S_{22}(\theta)=-1 \ .
\eeq
The S-matrix elements (\ref{eq.s}) do not have poles in the physical strip, 
therefore the two particles do not have bound states.

From this S-matrix we obtain the TBA equations for the ground state at 
cylinder circumference $L$, which take the form \cite{CastroAlvaredo:1999em,Dorey:2004qc}
\beq
\label{eq.tba}
\epsilon_a(\theta) + \sum_{b=1}^2 (K_{ab} * L_b)(\theta) =  m_a L \cosh(\theta)
 \, , \quad 
L_a(\theta) = \ln \left(1+\ee^{-\epsilon_a(\theta)}\right)\ , 
\eeq
with $a=1,2$, where 
$\epsilon_a(\theta)$ is the pseudo-energy function for the $a$-th particle and 
the kernels $K_{ab}=-\ii \frac{d}{d\theta}\ln S_{ab}(\theta)$ are 
\beq
K_{12}(\theta-\sigma)   = \frac{1}{\cosh(\theta)} = K_{21}(\theta+\sigma) \, , \quad 
K_{11}(\theta) = K_{22}(\theta)  =  0\ .
\eeq
The convolution here is defined as
\beq
(f*g)(\theta)=\int \frac{d\theta'}{2\pi} f(\theta-\theta')g(\theta')\ .
\eeq
It is convenient to use the dimensionless ground state energy, which is given by 
\beq
\label{eq.fl}
E_0(L)\frac{L}{2\pi}\equiv F(L)=-\frac{L}{4\pi^2}\sum_{a=1}^2  \int_{-\infty}^\infty d\theta\, m_a\cosh(\theta)\, 
L_a(\theta)+F_{\mathrm {bulk}}L^2\ .
\eeq
Here we had to add the term containing the bulk energy density $F_{\mathrm{bulk}}=\frac{1}{2\pi}E_{\mathrm{bulk}}$
to compensate the mismatch between the TBA and pCFT normalization of the ground-state energy.

In the following we introduce two real resonance parameters $\sigma_1$ and $\sigma_2$ , 
\beq
\label{eq.sigma}
\sigma=\sigma_1-\sigma_2 \, ,
\eeq
and two `left' and `right' masses, 
\beq
\label{eq.mua}
\mu_a=\frac{m_a \ee^{\sigma_a}}{2} \, ,\qquad \bar{\mu}_a=\frac{m_a \ee^{-\sigma_a}}{2} \, ,
\eeq
with $a=1,2$,
such that the shifted pseudo-energies
\beq
\hat{\epsilon}_a(\theta)=\epsilon_a(\theta-\sigma_a) 
\eeq
satisfy the TBA equations
\bea
\label{eq.tba1b}
\hat{\epsilon}_1(\theta) + (K * \hat{L}_2)(\theta) & = & 
L\left(\bar{\mu}_1 \ee^\theta +\mu_1 \ee^{-\theta}\right) \, , \\
\label{eq.tba2b}
\hat{\epsilon}_2(\theta) + (K * \hat{L}_1)(\theta) & = & L\left( \bar{\mu}_2\ee^\theta +\mu_2 \ee^{-\theta}\right)\ ,
\eea
with 
\beq
K(\theta)=\frac{1}{\cosh(\theta)} , \qquad 
\hat{L}_a(\theta)=L_a(\theta-\sigma_a) \ .
\eeq

Equation (\ref{eq.fl}), which gives the dimensionless ground state energy, takes the form 
\beq
\label{eq.fl2}
F(L) = -\frac{L}{4\pi^2} \sum_{a=1}^2 \int_{-\infty}^{+\infty} d\theta\,
  \left(\bar{\mu}_a \ee^\theta +\mu_a \ee^{-\theta}\right)   \hat{L}_a(\theta)+F_{\mathrm {bulk}}L^2  \ .
\eeq
The ground state energy is invariant under the following transformations:
\begin{enumerate}
\item 
Dynkin reflection: $\mu_1 \leftrightarrow \mu_2$,  $\bar{\mu}_1 \leftrightarrow \bar{\mu}_2$ ;
\item 
parity: $\mu_1 \leftrightarrow \bar{\mu}_1$, $\mu_2 \leftrightarrow \bar{\mu}_2$ ;
\item
scaling: $\mu_a \to \mu_a/\alpha$, $\bar{\mu}_a \to \alpha\bar{\mu}_a$, 
where $\alpha$ is any positive real number.
\end{enumerate}
The coefficient $F_{\mathrm{bulk}}$ can be calculated \cite{Hatsuda:2011ke}
by following the standard procedure for TBA systems:
\beq
\label{eq.Fbulk}
F_{\mathrm{bulk}}=\frac{1}{2\pi}(\mu_1\bar{\mu}_2+\bar{\mu}_1 \mu_2)\ .
\eeq
Dimensional analysis shows that the coefficients $F_n$, as functions of $\mu_a$ and $\bar{\mu}_a$,
have the scaling property
\beq
\label{eq.scalingF}
F_n(\alpha \mu_1, \alpha \mu_2, \alpha \bar{\mu}_1, \alpha \bar{\mu}_2) =
\alpha^{4n/5} F_n(\mu_1, \mu_2, \bar{\mu}_1,\bar{\mu}_2),\qquad \alpha>0\ .
\eeq

\subsection{Special cases}
\label{subsec.special}

In this TBA system, there are a few special cases at $\sigma=0$
in which it is possible to make definite statements about the 
relation between the pCFT and TBA parameters.
These cases provide  inputs and checks
for the exact mass-coupling relation  in the next section.

\paragraph{Single-mass cases ($\mathbf{m_1=0}$\ \ {\bf or}\ \ $\mathbf{m_2=0}$)}

In these cases the TBA of the HSG model coincides with the TBA of 
the (RSOS)$_3$ scattering theory, i.e., the unitary minimal model perturbed
by the primary field  of dimension $h=\bar{h}=3/5$;
$\mc{M}_{4,5} + \nu\phi_{1,3}$, $\nu < 0$.
Comparing the conformal perturbation series 
in this model and in the HSG model,  at second order we have
\beq
\brakett{\check{\Phi}\check{\Phi}} = \nu^2\brakett{\phi_{1,3}\phi_{1,3}}_{\mc{M}_{4,5}}\ ,
\eeq
 where $\check{\Phi} = \sum_{i,j}\nu_{ij}\Phi_{ij}$,
which gives 
\beq
p_2 \bar{p}_2 = \nu^2\ .
\eeq
At third order we have
\beq
\brakett{\check{\Phi}\check{\Phi}\check{\Phi}} =  -\nu^3 \brakett{\phi_{1,3}\phi_{1,3}\phi_{1,3}}_{\mc{M}_{4,5}}\ ,
\eeq
which gives
\beq
\label{eq.55}
p_3 \bar{p}_3= -\nu^3\ .
\eeq
The cases $m_1=0$ or $m_2=0$ thus correspond to $p_3 \bar{p}_3=p_2^{3/2} \bar{p}_2^{3/2}$, i.e.\ 
\beq
\frac{p_3}{p_2^{3/2}}=\frac{\bar{p}_3}{\bar{p}_2^{3/2}}=\pm 1\ .
\eeq
The  mass-coupling relation of the perturbed minimal models is known
\cite{Zamolodchikov:1995xk}.  In this case,
\beq
   \label{eq.MCrsos}
   \nu = - \kappa_3^{\rm RSOS} M^{4/5} \comma
\eeq
where $M$ is the mass of the massive particle and 
\beq
 \label{eq.kappa3}
  \kappa_3^{\rm RSOS} =
   \frac{1}{2(12\pi)^{1/5}} 
   \gamma^{\frac{1}{2}} \left(\frac{2}{5} \right) \gamma^{\frac{1}{2}}\left(\frac{4}{5} \right) \period
\eeq

\paragraph{Equal-mass case ($\mathbf{m_1=m_2}$)}
In the case $m_1=m_2=M$ 
the TBA for the ground state coincides with 
the TBA for the ground state of a non-unitary minimal model perturbed 
by the primary field of dimension $h= \bar{h} = 1/5$;
 $\mc{M}_{3,5} + \nu \phi_{1,3}$.  This  is equivalent
 to the perturbed diagonal  coset model $su(2)_1 \times su(2)_{-1/2}/su(2)_{1/2}$.
By comparing the conformal perturbation series in this model and in 
the HSG model one finds that 
\beq
\label{eq.ppp}
\brakett{\check{\Phi}\check{\Phi}\check{\Phi}} = 0
\eeq
has to hold in the HSG model. The reason for this is that 
in the perturbed $\mc{M}_{3,5}$ model $2-2 h=8/5$,
whereas in the HSG model $2-2h=4/5$, 
therefore odd terms in the perturbation series of the HSG model have to vanish.
From the OPEs (\ref{eq.ope}) one gets 
\begin{eqnarray}
\brakett{\check{\Phi}\check{\Phi}\check{\Phi}} & \propto & p_3\, \bar{p}_3\ ,
\label{eq.1}
\end{eqnarray}
thus (\ref{eq.ppp}) implies
\beq
p_3=\bar{p}_3=0\ .
\eeq
The  mass-coupling relation in this case reads \cite{Zamolodchikov:1995xk,Fateev:1993av}
\beq
   \nu =  \hat\kappa M^{8/5} \comma
\eeq
with 
\beq
\label{eq.kappahat}
 \hat\kappa^2 =
  - \frac{1}{(16\pi)^{2/5}} \left(\frac{5}{16} \right)^2
   \gamma \left(\frac{3}{5} \right)  \gamma \left(\frac{4}{5} \right)
    \gamma^{\frac{16}{5}}\left(\frac{1}{4} \right) \period
\eeq
Thus $\hat\kappa$ is purely imaginary.

\subsection{Form factors of the dimension $3/5$ operators}

The form factors are built from the minimal 2-particle form factors 
$F_{ab}(\theta_1,\theta_2)$ and polynomials in the variables $x_k^{\pm 1} =\ee^{\pm \theta_k}$.
The minimal form factors are \cite{CastroAlvaredo:2000em,CastroAlvaredo:2000nk}
\bea
F_{ab}(\theta_1,\theta_2) & = &
f_{ab}(\theta_1-\theta_2),\qquad \mathrm{if}\quad a\not= b,\\
F_{ab}(\theta_1,\theta_2) & = &
\frac{-\ii f_{aa}(\theta_1-\theta_2)}{2\pi(x_1+x_2)},\qquad \mathrm{if}\quad a=b,
\eea
where
\beq
f_{11}(\theta)=f_{22}(\theta)=-\ii\sinh\frac{\theta}{2}
\eeq 
and 
\beq
f_{12}(\theta)={\cal G}(\theta-\ii\pi),\qquad\quad
f_{21}(\theta)={\cal G}(\ii\pi-\theta)
\eeq 
with 
\beq 
{\cal G}(\theta)=2^{-\frac{1}{4}}\exp\left\{\frac{\sigma}{4}-\frac{G}{\pi}+\frac{\theta}{4}
-\int_0^\infty\frac{{\rm d}t}{t}\,\frac{\sin^2\frac{(\theta+\sigma)t}{2\pi}}
{\sinh t\cosh(t/2)}\right\}.
\eeq
Here  $G=0.91597\dots$ is the Catalan constant. Note that for many calculations we do not need
this explicit integral representation and it is sufficient to use the relation
\beq
{\cal G}(\theta){\cal G}(\theta-\ii\pi)=\frac{1}{1+\ii{\rm e}^{-\sigma-\theta}}.
\eeq

A general $n$-particle form factor corresponding to a local operator $X$ takes the form
\beq
\label{eq.FX}
{\cal F}^X_{a_1\dots a_n}(\theta_1,\dots,\theta_n)=
\left[\prod_{i<j}F_{a_ia_j}(\theta_i,\theta_j)\right]\,Q^X_{a_1\dots a_n}(x_1,\dots,x_n).
\eeq 
For $X=\Theta$, the trace of the energy-momentum (EM) tensor, 
the 2-particle solution is characterized by
\beq
Q^\Theta_{11}=\ii m_1^2(x_1+x_2),\qquad\quad
Q^\Theta_{22}=\ii m_2^2(x_1+x_2),
\eeq 
and the 4-particle form factor corresponds to
\beq
\label{eq.QTh}
Q^\Theta_{1122}(x_1,x_2,x_3,x_4)=-2\ee^{-\sigma} P^2 x_3x_4,
\eeq 
where $P^2=P^+P^-$ is the square of the total momentum
\beq
P^\mu=m_1\hat P^\mu_{(1)}+m_2\hat P^\mu_{(2)}.
\eeq 
Here for later purposes we introduced the notation $\hat P^\mu_{(a)}$, the 
coefficient of $m_{a}$ in the total momentum,
with $\mu=+,-$;  $a=1,2$. 

It is clear that all form factors of the trace $\Theta$ are
proportional to $P^2$ and take the form
\beq
\label{eq.thff}
Q^\Theta_{a_1\dots a_n}=P^2\,q_{a_1\dots a_n}.
\eeq 
$q_{a_1\dots a_n}$ itself also satisfies almost all the requirements coming from
the form factor axioms, with the exception of the $n=2$ case where it would lead
to a singular form factor. This singularity is cancelled by the prefactor $P^2$.
This cancellation also happens if we consider
various parts of $P^2$ separately. This way we can introduce the local operators $A,B,C,D$,
whose form factors (satisfying all the requirements including the cancellation of the above
mentioned pole in the 2-particle case) are defined by
\bea
Q^A_{a_1\dots a_n} & = & \hat P^+_{(1)}\hat P^-_{(1)}\,q_{a_1\dots a_n},
\label{eq.QabcdA} \\
Q^B_{a_1\dots a_n} & = & \hat P^+_{(2)}\hat P^-_{(2)}\,q_{a_1\dots a_n},\\
Q^C_{a_1\dots a_n} & = & (\hat P^+_{(1)}\hat P^-_{(2)}+\hat P^+_{(2)}\hat P^-_{(1)})
\,q_{a_1\dots a_n},\\
Q^D_{a_1\dots a_n} & = & (\hat P^+_{(1)}\hat P^-_{(2)}-\hat P^+_{(2)}\hat P^-_{(1)})
\,q_{a_1\dots a_n}
\label{eq.QabcdD} .
\eea
It is clear from (\ref{eq.thff}) that
\beq
\label{eq.ThetaABCD}
\Theta = m_1^2 A + m_2^2 B + m_1m_2 C.
\eeq 
In the 4-particle example (see Appendix \ref{sect.formfactor}, where all higher 
form factors can be found), 
\beq
q_{1122}(x_1,x_2,x_3,x_4)=-2\ee^{-\sigma} x_3x_4.
\eeq 

For later purposes we now calculate the 1-particle and 2-particle diagonal form factors
\beq
\label{eq.diagFF}
{\cal F}_{ab}^{X(s)}(\theta)
=\lim_{\epsilon\to0}{\cal F}_{ab}^{X}(\theta+\ii\pi+\epsilon,\theta),\qquad
{\cal F}_{abab}^{X(s)}(\theta_1,\theta_2) =  \lim_{\epsilon\to 0}
{\cal F}_{abab}^X(\theta_1+\ii\pi+\epsilon,\theta_2+\ii\pi+\epsilon,\theta_1,\theta_2).
\eeq
Here the superscript $(s)$ refers to the symmetric evaluation. We find
\beq
 \label{eq.2FF}
{\cal F}_{11}^{A(s)}(\theta)={\cal F}_{22}^{B(s)}(\theta)=\frac{1}{2\pi}.
\eeq
All other 1-particle diagonal form factors vanish.

For the 2-particle case, from the explicit form of  $q_{a_1...a_n}$ calculated in 
\cite{CastroAlvaredo:2000em,CastroAlvaredo:2000nk}, we obtain
\beq
\label{eq.4FF}
{\cal F}_{1212}^{X(s)}(\theta_1,\theta_2) 
= \left.-\frac{\ii}{4\pi^2}\,\frac{\partial S_{12}}{\partial\theta}(\theta)\{1,1,
2\cosh\theta,2\sinh\theta\}\right\vert_{\theta=\theta_1-\theta_2}
\eeq
for the operators $X=\{A,B,C,D\}$ respectively.

The form factors of $A,B,C,D$ are obtained from those of $\Theta$
by replacing  the momenta $P^\mu$  with $m_a \hat{P}_{(a)}^{\mu}$. 
From this similarity, one may expect that these operators have the same
dimension $3/5$ as $\Theta \sim \mathcal{L}_{\rm pert}$. 
This is checked numerically below.

\subsection{Numerical check of the dimension}

To find the dimension of  the operators $A,B,C,D$, 
we consider the two-point functions,
\beq
  \langle \calO_i(r) \calO_j(0) \rangle = \sum_k C_{ij}^{\ \,  k} r^{2h_k - 2 h_i - 2h_j}
    \langle \calO_k(0) \rangle  + \cdots \comma
\eeq
for small $r$. The constants $C_{ij}^{\ \, k}$ are the three-point couplings.
Since the $su(3)_2/u(1)^2$ HSG model is unitary, the dominant contribution 
comes from the identity $ I = \calO_{k = 0}$ with $h_0 =0$ on the right-hand side, 
and thus 
$ \langle \calO_i(r) \calO_i(0) \rangle \sim r^{-4h_i}$  for small $r$ if $C_{ii}^{\ \, 0} \neq 0$. 
Furthermore,  the two-point functions are evaluated 
by the form factors through the expansion,
\beq
   \langle \calO (r) \calO (0) \rangle
   = \sum_{n=0}^{\infty} \sum_{a_1,\dots, a_n} 
   \int_{-\infty}^\infty \frac{d\theta_1 \dots d\theta_n}{n!}
   \exp\Bigl( -r \sum_{i=1}^n  m_{a_i} \cosh \theta_i \Bigr)
   \big\vert \calF^\calO_{a_1 \dots a_n} (\theta_1,\dots, \theta_n)\big\vert^2
   \period
\eeq

We have performed the multi-dimensional integrals 
for $ A $ and $ (C+D)/2$. 
For simplicity, we have set $m_1 = m_2 =: m $ and 
$\sigma$ (resonance parameter) $= 0$. 
For the relevant form factors, see Appendix \ref{sect.formfactor}.
Alternatively, in such a  case with   $m_1 = m_2 $,
the explicit forms of the form factors of $\Theta$  are found in 
\cite{CastroAlvaredo:2000em,CastroAlvaredo:2000nk} up to the 8-particle ones.
Similarly to the cases of (\ref{eq.2FF}), (\ref{eq.4FF}), 
one can convert these results to the form factors of $A,B,C,D$ via
(\ref{eq.FX}) and (\ref{eq.thff})-(\ref{eq.QabcdD}).

\begin{figure}[t]
\begin{center}
\begin{tabular}{cc}
\resizebox{90mm}{!}{\includegraphics{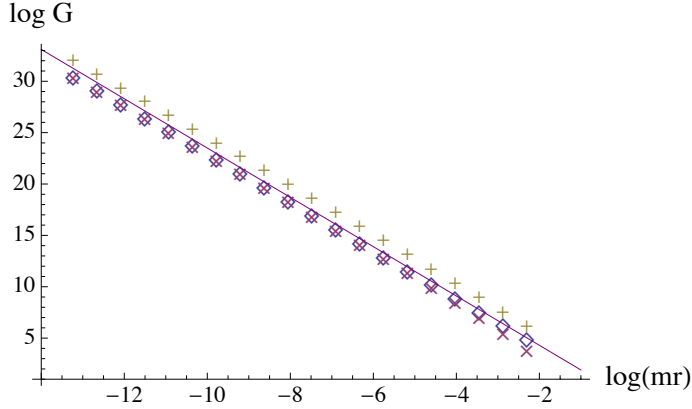}}
\vspace{-3mm}
\end{tabular}
\end{center}
\caption{Plots of $ \log \langle \calO(r) \calO(0) \rangle =: \log G $ 
for $O=A$ ($\diamond$), $(C+D)/2$ ($\times$) and $\Theta$ ($+$). 
We have set $m_1=m_2=:m$ and $\sigma = 0$.
The solid line represents $-(12/5) \log(mr) +$(const.). }
\label{fig:2ptFn} 
\end{figure}

Figure \ref{fig:2ptFn} shows plots of $\log (mr)$ versus $ \log \langle \calO(r) \calO(0) \rangle
=: \log G $ for $O=A$ ($\diamond$)  and $(C+D)/2$ ($\times$).
The contributions of the $n$-particle form factors are included up to $n=6$.
(For $(C+D)/2$, the 2-particle contribution vanishes.)   We have omitted 
the constants for $n=0$, which are irrelevant to small $r$ behavior. These vacuum
expectation values are obtained in the next section. 
For reference,  a similar plot for $\Theta$ 
($+$), and a plot of $-(12/5) \log (mr) +$(const.) (solid line) are shown.
We observe that all of these  data scale approximately as
$-(12/5) \log (mr)$, which is consistent with $h=3/5$.
The results for  $B, (C-D)/2$ follow from the symmetry $m_1 \leftrightarrow m_2$.

Thus, the operators $A,B,C,D$ may form a basis of the IR counterpart
of $\Phi_{ij}$. 
The way to define these operators via form factors,  
$P^\mu \to m_a \hat{P}^\mu_{(a)}$, was simple. Yet, applying 
a similar replacement to the EM tensor, 
we can obtain additional conserved currents on the IR side,
which will play an important role in the following discussion.

\section{Analytical mass-coupling relation}
\label{sec.AnMCrel}

Based on the results from the UV and the IR side in the previous sections,
we derive the exact mass-coupling relation in this section.
First, using the formulas for the response of the physical masses and the S-matrix
under the change of the couplings, we find explicit relation between the UV operators 
$\Psi_i $ in (\ref{eq.Consv2}) 
and the IR ones $A,B,C,D$. This enables us to find the form factors of $\Psi_i$.
Expressing the conservation laws in Section \ref{sec.HSGpCFT} 
by the perturbing operators $\Psi_i$,
we then show that the ratio $\mu_1/\mu_2$ depends only on $\lambda_i$, not on $\bar\lambda_i$
(`partial factorization'), which also simplifies the IR expression of $\Psi_i$. 
As mentioned above, using the  partial momenta $P_{(a)}^\mu$
we obtain conserved currents in terms of the IR variables. 
They are identified with the UV currents through the IR expression 
of $\Psi_i$.
By comparing their commutation relations on the UV and the IR side, the perturbing
operators $\Phi_{ij}$ are expressed by the IR operators.
With the help of  the generalized $\Theta$ sum rule for the above conserved 
current and $\bar\Psi_j$, the free energy Ward identity relates
the vacuum expectation values (VEV) of $\Phi_{ij}$ and
the derivatives of the free energy with respect to the couplings.
From this relation,  $\mu_a$ themselves 
are found to be functions of $\lambda_i$ only (`complete factorization').
 Applying again the generalized $\Theta$ sum rule to $\Phi_{ij}$, 
the Ward identity for $\Phi_{ij}$ 
yields a differential equation for their vacuum expectation values. 
This is further translated into a differential equation for the mass-coupling relation,
which is solved by hypergeometric functions.

To begin with, let us recall the form of the perturbing Lagrangian on the UV side,
\begin{equation}
\mc{L}_{\rm pert}(z,\bar z)=\lambda_1\bar{\lambda}_1\Phi_{11}(z,\bar z)+
\lambda_1\bar{\lambda}_2\Phi_{12}(z,\bar z)+
\lambda_2\bar{\lambda}_1\Phi_{21}(z,\bar z)+
\lambda_2\bar{\lambda}_2\Phi_{22}(z,\bar z) .
\end{equation}

\subsection{Exact VEVs and relations from changing the couplings}

Here we collect all available pieces of information about the coupling dependence of the
problem. First of all we establish that because all perturbing operators are of dimension
$(3/5,3/5)$, the trace of the EM tensor is given by
\beq
\label{eq.ThetaLpert}
\Theta=-\frac{4}{5}{\cal L}_{\rm pert}.
\eeq 
Further, the VEV of the EM tensor must be of the form
\beq
\langle T_{\mu\nu}\rangle= \varepsilon\eta_{\mu\nu},
\eeq 
where $\eta_{\mu\nu}$ is the $1+1$ dimensional Minkowski metric. The bulk energy
density is known from TBA as  in (\ref{eq.Fbulk}), which we denote by 
\beq
\varepsilon=\langle T_{00}\rangle =\frac{m_1m_2}{2}\cosh\sigma. 
\eeq 
Thus
\beq
\label{eq.thetavev}
\langle \Theta\rangle=2\varepsilon=m_1m_2 \cosh\sigma. 
\eeq 
We can also calculate the free energy density ${\cal F}$. First we have to calculate the
partition function ${\cal Z}$ in finite 2-volume ${\cal V}$ and then take the limit 
\beq
{\cal F}=-\lim_{{\cal V}\to\infty}\frac{1}{{\cal V}}\ln{\cal Z}.
\eeq 
From this definition it is easy to see that a small change of the couplings leads to the
relations
\beq
\frac{\partial{\cal F}}{\partial\lambda_i}=-\langle\Psi_i\rangle \, ,\qquad
\Psi_i = \frac{\partial{\cal L}_{\rm pert}}{\partial\lambda_i}=
\bar{\lambda}_1\Phi_{i1}+\bar{\lambda}_2\Phi_{i2} \,  .
\label{eq.firstder}
\eeq
In the following we refrain from writing out explicitly analogous equations for the bar
variables if it is obviously true with the left/right replacement. 

Since ${\cal F}$ is of mass dimension 2, from dimensional analysis we get
\beq
\sum_{i=1}^2\,\lambda_i\frac{\partial{\cal F}}{\partial\lambda_i}=
\frac{5}{2}\,{\cal F}.
\eeq 
 Thus $\frac{5}{2}{\cal F}=\frac{5}{4}\langle\Theta\rangle $ and hence
\beq
\label{eq.Fep}
{\cal F}=\varepsilon=\frac{m_1m_2}{2}\cosh\sigma, 
\eeq 
as anticipated.

The result of infinitesimal changes of the couplings can be expressed in terms of
the matrix elements of the operators $\Psi_i$, $\bar{\Psi}_j$ \cite{Delfino:1996xp}.
For example, the change
of the particle mass is given by 
\beq
\label{eq.FFPsi}
\frac{\partial m_a^2}{\partial \lambda_i}=-4\pi\langle 0,a\vert \Psi_i(0,0)\vert
0,a\rangle=-4\pi {\cal F}_{aa}^{\Psi_i(s)}(0) , 
\eeq 
while the change of the scattering matrix is given by the formula
\bea
\label{eq.FdS}
4\pi^2 \ii{\cal F}_{abab}^{\Psi_i(s)}(\theta_1,\theta_2) & = & -\left(
\frac{1}{2}\frac{\partial m_a^2}{\partial \lambda_i}+
\frac{1}{2}\frac{\partial m_b^2}{\partial \lambda_i}+
\cosh\theta\frac{\partial (m_a m_b)}{\partial \lambda_i}\right)
\frac{\partial S_{ab}(\theta)}{\partial\theta} \nonumber\\ 
&& +m_am_b\sinh\theta
\frac{\partial S_{ab}(\theta)}{\partial\lambda_i},
\eea 
where $\theta=\theta_1-\theta_2$, and  similar ones for the bar variables.

\subsection{Relations among the local operators}
\label{subsec.relLO}

It is very natural to assume that the local operators $\Psi_i$, $\bar{\Psi}_j$
related to the pCFT Lagrangian and the operators $A$, $B$, $C$, $D$ defined on
the form factor side form the same operator basis. Their relation can be 
written as
\bea
\label{eq.PsiABCD}
\Psi_i & = & X^A_iA+X^B_iB+X^C_iC+X^D_iD ,
\eea
with some coefficients $X^A_i$ etc. and similarly with $\bar{X}^A_j$ for $\bar{\Psi}_j$.
The coefficients are not all independent since they
have to satisfy  
the relations which follow from (\ref{eq.ThetaABCD}) and (\ref{eq.ThetaLpert}),
\beq
\sum_{i=1}^2\lambda_iX^A_i=-\frac{5}{4}m_1^2,\quad
\sum_{i=1}^2\lambda_iX^B_i=-\frac{5}{4}m_2^2,\quad
\sum_{i=1}^2\lambda_iX^C_i=-\frac{5}{4}m_1m_2,\quad
\sum_{i=1}^2\lambda_iX^D_i=0 \, .
\eeq 
Taking into consideration these relations, from the mass dependence of the 
VEV of $\Theta$ (\ref{eq.thetavev}) we 
have
\beq
\label{eq.ABCvev}
\langle A\rangle=\langle B\rangle=0,\qquad\quad
\langle C\rangle=\cosh\sigma ,
\eeq 
and this leads to
\beq
-\frac{\partial{\cal F}}{\partial\lambda_i}=X^C_i\cosh\sigma+X^D_i\langle D\rangle.
\label{eq.free}
\eeq
From the mass relation 
(\ref{eq.FFPsi}) and the form factors (\ref{eq.2FF})
we obtain 
\beq
X^A_i=-\frac{1}{2}\frac{\partial m_1^2}{\partial\lambda_i},\qquad
X^B_i=-\frac{1}{2}\frac{\partial m_2^2}{\partial\lambda_i}.
\eeq 
Finally, from the S-matrix formula (\ref{eq.FdS}) and the form factors (\ref{eq.4FF}) 
we can read off
\bea
&& X^C_i=-\frac{1}{2}\frac{\partial (m_1m_2)}{\partial\lambda_i}\ ,\qquad\quad 
X^D_i=\frac{1}{2}m_1m_2\frac{\partial \sigma}{\partial\lambda_i} .
\eea
Comparing to (\ref{eq.free}) with $\mc{F}$ in (\ref{eq.Fep}) we see that they are consistent if
\beq
\label{eq.Dvev}
\langle D\rangle=-\sinh\sigma.
\eeq
Having found the coefficients we can now write down the complete expression for
the perturbing operators $\Psi_i, \bar\Psi_j$ in terms of the bootstrap ones.

\subsection{Relations from conserved spin 1 charges: factorization of mass ratios}
\label{subsec.spin1}

Given the IR expression of the perturbing operators, we can derive
non-trivial relations from the conserved currents. To see this, we first recall that
the  form factors $\mc{F}^{\Psi_i}$ take the form (\ref{eq.FX}). 
Substituting (\ref{eq.QabcdA})-(\ref{eq.QabcdD}) and factoring out the minimal 2-particle 
form factors and $q_{a_1...q_n}$,
we are left with the proportionality coefficient 
\beq
\begin{split}
&-(\partial_i\ln m_1)P^+_{(1)}P^-_{(1)}-
(\partial_i\ln m_2)P^+_{(2)}P^-_{(2)}\\
&-\frac{1}{2}(\partial_i\ln m_1+\partial_i \ln m_2+\partial_i\sigma)P^+_{(2)}P^-_{(1)}
-\frac{1}{2}(\partial_i\ln m_1+\partial_i \ln m_2-\partial_i\sigma)P^+_{(1)}P^-_{(2)},
\end{split}
\label{eq.coeff}
\eeq
where $\partial_i=\partial/\partial\lambda_i$.
There is an analogous formula for $\bar{\Psi}_j$. Note that this formula is written
in terms of $P^\mu_{(1,2)}$, the $1,2$ parts of the full momentum:
\beq
P^\mu_{(a)}=m_a\hat P^\mu_{(a)}\ ,\qquad\quad P^\mu=P^\mu_{(1)}+P^\mu_{(2)}\ .
\eeq

For our purposes we now write the conservation laws 
(\ref{eq.EMconsv}) and (\ref{eq.Consv2}) in the form
\begin{equation}
\partial \Psi_i=\bar\partial \tau_i\ ,
\end{equation} 
where the local operators $\tau_i$ are some linear combinations of the $L^{(i)}$s.
The Minkowski version of these spin-1 
conservation laws in the language of form factors imply
that the form factors of $\Psi_i$ are proportional to the $^+$ light-cone component
of the total momentum:
\begin{equation}
{\cal F}^{\Psi_i}=P^+\,f_i\ ,\qquad\quad {\cal F}^{\tau_i}=P^-\,f_i\ .
\end{equation} 

The requirement that (\ref{eq.coeff}) is proportional to $P^+$, 
though rather obvious from the UV point of view,  leads to the two equivalent
relations
\bea
2\partial_i\ln m_1 & = & \partial_i\ln m_1+\partial_i\ln m_2+\partial_i\sigma,
\eea
implying
\begin{equation}
\partial_i\ln\left(\frac{m_1}{m_2}{\rm e}^{-\sigma}\right)= 
\partial_i\ln\left(\frac{\bar\mu_1}{\bar\mu_2}\right)=0. 
\end{equation} 
Thus the chiral mass ratio $\mu_1/\mu_2$ only depends on $\lambda_i$ and similarly
$\bar\mu_1/\bar\mu_2$ only depends on $\bar\lambda_i$, showing the `partial factorization'. 
With this simplification the proportionality coefficients are also simplified, to give 
\beq
  \mc{F}^{\Psi_i} \propto
-(\partial_i\ln m_1)P^+P^-_{(1)}-(\partial_i\ln m_2)P^+P^-_{(2)}
\label{eq.coeff2}
\eeq
for $\Psi_i$, and there is an analogous relation for $\bar\Psi_j$.

To prove the full factorization we have to study further properties of the conserved currents.

\subsection{Relations from conserved tensor currents}

Next, we consider conserved tensor currents.
Using the \lq\lq scalarized'' form factors of $\Theta$, we can define, via their
form factors, the tensor operators $X^{\mu\nu}_{(a)(b)}$. The corresponding
form factors are
\beq
Q^{X^{\mu\nu}_{(a)(b)}}_{a_1\dots a_n}=P^\mu_{(a)}P^\nu_{(b)}q_{a_1\dots a_n} ,
\eeq 
and they are local operators since the two momentum factors cancel the unwanted
double pole from the 2-particle form factors. Since all operators we consider here
are proportional to the \lq\lq scalarized'' form factors $q_{a_1\dots a_n}$, we will
use the simplified notation\footnote{$X^{+-}_{(a)(b)}$ are denoted by $X_{ab}$ in \cite{Bajnok:2015eng}.}
\beq
X^{\mu\nu}_{(a)(b)}\sim P^\mu_{(a)}P^\nu_{(b)},\qquad\quad
\Theta\sim P^2.
\eeq
The scalar operators we introduced earlier are given in this new notation as
\beq
m_1^2A=X^{+-}_{(1)(1)},\qquad
\quad
m_2^2B=X^{+-}_{(2)(2)} ,
\eeq
and
\beq
m_1m_2C=X^{+-}_{(1)(2)}+X^{+-}_{(2)(1)},\qquad\quad
m_1m_2D=X^{+-}_{(1)(2)}-X^{+-}_{(2)(1)}.
\eeq
For later use we list here the vacuum expectation values in the new notation
\beq
\langle X^{+-}_{(1)(1)}\rangle=\langle X^{+-}_{(2)(2)}\rangle=0,\qquad\quad
\label{eq.vev1}
\eeq
\beq
\langle X^{+-}_{(1)(2)}\rangle=\frac{1}{2}m_1m_2{\rm e}^{-\sigma}
=2\mu_2\bar\mu_1,\qquad\quad
\langle X^{+-}_{(2)(1)}\rangle=\frac{1}{2}m_1m_2{\rm e}^\sigma
=2\mu_1\bar\mu_2.
\label{eq.vev2}
\eeq
We also introduce
\beq
Y^{\mu\nu}_{(a)}=\sum_b X^{\mu\nu}_{(b)(a)}\sim P^\mu P^\nu_{(a)},
\eeq
where
\beq
P^\mu=\sum_a P^\mu_{(a)}
\eeq
is the total momentum, and
\beq 
Z^{\mu\nu}=\sum_a Y^{\mu\nu}_{(a)}\sim P^\mu P^\nu.
\eeq
The energy-momentum tensor in this notation is
\beq
T^{\mu\nu}=-\epsilon^{\mu\alpha}\epsilon^{\nu\beta} Z_{\alpha\beta}\sim
-\epsilon^{\mu\alpha}\epsilon^{\nu\beta} P_\alpha P_\beta.
\eeq
Its conservation is obvious in this representation. We can now define further
conserved tensor currents by
\beq
I^{\mu[\nu]}_{(a)}=-\epsilon^{\mu\alpha}\epsilon^{\nu\beta} Y_{\alpha\beta(a)}\sim
-\epsilon^{\mu\alpha}\epsilon^{\nu\beta} P_\alpha P_{\beta (a)}.
\eeq
These are also obviously conserved in their first indices:
\beq
\partial_\mu I^{\mu[\nu]}_{(a)}=0.
\eeq
We put the second tensor index to square brackets to indicate that it is part of 
the \lq\lq name'' of the conserved current (together with the particle subscript $(a)$).
There are altogether four conserved currents, but two combinations of them 
are not new, because of the relation
\beq 
\sum_a I^{\mu[\nu]}_{(a)}=T^{\mu\nu}.
\eeq

The corresponding conserved charges are given by
\beq
Q^{[\nu]}_{(a)}=\int {\rm d}x I^{\,\,\,[\nu]}_{0(a)}(x,t).
\eeq
These act diagonally on multi-particle states
\beq
Q^{[\nu]}_{(a)}\vert \theta_1,a_1;\dots;\theta_n,a_n\rangle=
P^\nu_{(a)}\vert \theta_1,a_1;\dots;\theta_n,a_n\rangle.
\eeq
The above eigenvalues can be obtained by first considering one-particle
states, where the eigenvalues can be calculated directly from the two-particle form factors,
and then using additivity for multi-particle states. The latter property of the conserved
charges follows from the fact that they are given as space integrals of local currents.
The physical meaning of the conserved charges is thus rather trivial: they just express
the separate conservation of the two parts of the total momentum corresponding to each particle
type. These parts are trivially conserved since the particle momenta are not changed in a 
scattering process since the scattering is diagonal.

The algebra of the conserved charges is Abelian, but we can obtain useful information by 
considering the action of the charges on the local current components. We find
\beq
[Q^{[\nu]}_{(a)},I^{\mu[\rho]}_{(b)}]\sim 
P^\nu_{(a)}\epsilon^{\mu\alpha}\epsilon^{\rho\beta}P_\alpha P_{\beta(b)}\sim 
\ii\epsilon^{\mu\alpha}\partial_\alpha\epsilon^{\rho\beta}X^\nu_{(a)(b)\beta}.
\eeq
Here we used the fact that the form factors of the derivative of a local operator are
proportional to the original form factor multiplied by the total momentum.
The commutator formula becomes more transparent if we specify some of the tensor indices:
\beq
[Q^{[+]}_{(a)},I^{\mu[-]}_{(b)}]=-\ii\epsilon^{\mu\alpha}\partial_\alpha X^{+-}_{(a)(b)},
\label{eq.comm1}
\eeq
\beq
[Q^{[-]}_{(b)},I^{\mu[+]}_{(a)}]=\ii\epsilon^{\mu\alpha}\partial_\alpha X^{+-}_{(a)(b)}.
\label{eq.comm2}
\eeq

Now we identify the conserved currents and charges on the pCFT side. We already established 
in (\ref{eq.Consv2}) that
(after Wick-rotating the pCFT formulas to Minkowski space)
\beq
\partial_+ J^++\partial_-J^-=0,
\eeq
where
\beq
J^-=-2L^{(1)}-2\kappa L^{(3)},\quad 
\kappa=\frac{\sqrt{5}}{4}\frac{\lambda_1}{\lambda_2},\quad 
J^+=v_i\Psi_i,\quad v_1=\pi\lambda_1,\quad v_2=\frac{\pi}{3}\frac{\lambda_1^2}{\lambda_2}.
\eeq
We will denote the corresponding charge by $Q$.
Since we already expressed the scalar operators $\Psi_i$ in terms of the 
$X^{+-}_{(a)(b)}$ basis on the IR side, we can write
\beq
J^+=-v_i(\partial_i\ln m_a)I^{+[-]}_{(a)}.
\eeq
Because of Lorentz covariance, the same linear combination has to appear for the other
tensor component as well and we can write:
\beq
J^\mu=-v_i(\partial_i\ln m_a)I^{\mu[-]}_{(a)}.
\eeq
Analogous formulas exist for the bar variables, so we can summarize the relation  
between the current components in the UV and IR bases as
\beq
J^\mu=-k_a I^{\mu[-]}_{(a)},\qquad\quad k_a=v_i(\partial_i\ln m_a),
\eeq
\beq
\bar J^\mu=-\bar k_a I^{\mu[+]}_{(a)},\qquad\quad \bar k_a=\bar v_j(\bar \partial_j\ln m_a).
\eeq
Using this identification and the commutation relations (\ref{eq.comm1}) and (\ref{eq.comm2})
we get
\beq
\label{eq.QJ}
[Q,\bar J^\mu]=-[\bar Q,J^\mu]=\ii\epsilon^{\mu\alpha}\partial_\alpha \Omega,
\eeq
where
\beq
\Omega=\bar k_a k_b X^{+-}_{(a)(b)}.
\label{eq.Omega1}
\eeq
These are used to express the perturbing operators $\Phi_{ij}$ in the IR basis.

\subsection{Relation between the UV and IR bases}

So far we are able to give the UV scalars $\Psi_i$ and $\bar\Psi_j$ in terms of the IR
scalars $X^{+-}_{(a)(b)}$. Only three linear combinations are independent, because of the
relation $\lambda_i\Psi_i=\bar\lambda_j\bar\Psi_j$.
Here we will determine the remaining coefficients ${\cal N}^{ab}_{ij}$ in the relation 
\beq
\label{eq.PhiNX}
\Phi_{ij}=\phi_i\bar\phi_j={\cal N}^{ab}_{ij} X^{+-}_{(a)(b)}.
\eeq
We start from
\beq
\label{eq.QJ2}
[Q,\bar J^-]=-2\ii\partial_+\Omega=-2\partial\Omega.
\eeq
(Here the last equality comes from continuing back the formula to pCFT conventions.)
At the leading order we  have
\beq
[Q,\bar J^-]= -\pi\oint\frac{{\rm d}w}{2\pi \ii}\left\{J^-(w)\bar v_j\bar\Psi_j(z,\bar z)\right\}
  =\partial\left(\frac{5}{2}
v_i\bar v_j\phi_i\bar\phi_j\right) ,
\eeq
where the short distance expansion formulas
summarized in Section \ref{subsec.MMrep} 
were used.
By dimensional analysis  explained in Section \ref{sec.HSGpCFT}, we can convince
ourselves that this leading order formula is actually exact in conform perturbation theory. 
Comparing this to (\ref{eq.QJ2}) 
and  (\ref{eq.Omega1}),
\beq
\Omega=-\frac{5}{4}v_i\bar v_j\phi_i\bar\phi_j=-\frac{5}{4}v_i\bar v_j\Phi_{ij} ,
\eeq
in terms of UV fields, and 
\beq
v_i(\partial_i\ln m_b)\bar v_j(\bar\partial_j\ln m_a)=-\frac{5}{4}v_i\bar v_j{\cal N}^{ab}_{ij}.
\label{eq.omega}
\eeq
Using also the relations we found earlier,
\beq
\bar\Psi_j=\lambda_i{\cal N}^{ab}_{ij} X^{+-}_{(a)(b)}=-(\bar\partial_j\ln m_a)Y^{-+}_{(a)}=
-\sum_b(\bar\partial_j\ln m_a)X^{+-}_{(a)(b)}
\eeq
we have
\beq
\lambda_i{\cal N}^{ab}_{ij}=-\bar\partial_j\ln m_a.
\eeq
Similarly we have
\beq
{\cal N}^{ab}_{ij}\bar\lambda_j=-\partial_i\ln m_b.
\eeq
These two relations, together with (\ref{eq.omega}) imply
\beq
{\cal N}^{ab}_{ij}=-\frac{4}{5}(\partial_i\ln m_b)(\bar\partial_j\ln m_a).
\label{eq.Nabij}
\eeq
Thus we completely identified the four UV scalars $\Phi_{ij}$ in terms of the IR scalars
$X^{+-}_{(a)(b)}$.

\subsection{Free energy Ward identity}

From the IR expression of $\Phi_{ij}$, we
can  prove the full factorization. For this purpose, we also need
the free energy Ward identity, which is discussed below.

Using the vacuum expectation values (\ref{eq.vev1}), (\ref{eq.vev2}) and (\ref{eq.Nabij}) we can 
calculate the vacuum expectation values of $\Phi_{ij}$:
\beq
\langle\Phi_{ij}\rangle=
-\frac{2}{5}\mu_1\bar\mu_2(\partial_i\ln\mu_1\bar\mu_1)
(\bar\partial_j\ln\mu_2\bar\mu_2)
-\frac{2}{5}\mu_2\bar\mu_1(\partial_i\ln\mu_2\bar\mu_2)
(\bar\partial_j\ln\mu_1\bar\mu_1).
\label{eq.Phiijvev}
\eeq
Here we used the chiral mass  parameters defined in 
(\ref{eq.mua}).
The free energy density can be similarly written as a sum of two chirally factorized terms:
\beq
{\cal F}=\mu_1\bar\mu_2+\mu_2\bar\mu_1.
\label{eq.chiralF}
\eeq
Using (\ref{eq.firstder}) and taking a second derivative with respect to the couplings
we can derive the following Ward identity:
\beq
\partial_i\bar\partial_j{\cal F}=-\langle\Phi_{ij}\rangle-\int{\rm d}^2x
\langle\Psi_i(x)\bar\Psi_j(0)\rangle_c.
\eeq
Here the subscript $_c$ means, as before, connected correlation function.

We can calculate the integral of the two-point correlation functions by using the generalized
$\Theta$ sum rule, which is derived in Appendix \ref{app.d}. 
First we calculate
\beq
\int{\rm d}^2x \langle J^+(x)\bar\Psi_j(0)\rangle_c=
v_i\int{\rm d}^2x \langle\Psi_i(x)\bar\Psi_j(0)\rangle_c,
\eeq
for which we need the short distance  expansion
\beq
\begin{split}
&J^-(z,\bar z)\bar\Psi_j(0,0)
 \approx-\frac{3}{2\pi z^2}v_i\phi_i(0)\bar\phi_j(0).
\end{split}
\eeq
Using this in the generalized $\Theta$ sum rule we obtain
\beq
v_i\int{\rm d}^2x \langle\Psi_i(x)\bar\Psi_j(0)\rangle_c
=\frac{3}{2}v_i\langle\Phi_{ij}\rangle.
\label{eq.sum1}
\eeq
In our case the original $\Theta$ sum rule, 
where $v_i$ are replaced by $\lambda_i$, 
  gives
\beq
\lambda_i\int{\rm d}^2x \langle\Psi_i(x)\bar\Psi_j(0)\rangle_c=
-\frac{5}{4}\int{\rm d}^2x \langle \Theta(x)\bar\Psi_j(0)\rangle_c
=
\frac{3}{2}\lambda_i\langle\Phi_{ij}\rangle.
\eeq
The last two relations together imply
\beq
\int{\rm d}^2x \langle\Psi_i(x)\bar\Psi_j(0)\rangle_c
=\frac{3}{2}\langle\Phi_{ij}\rangle \, , 
\eeq
and putting this result into the free energy sum rule leads to the simple relation
\beq
\partial_i\bar\partial_j{\cal F}=-\frac{5}{2}\langle\Phi_{ij}\rangle.
\label{eq.simple}
\eeq

\subsection{Proof of complete factorization}
\label{subsec.compfactr}

The partial factorization we already established in subsection \ref{subsec.spin1}
allows the following parametrization:
\beq
\mu_1=\mu,\qquad \bar\mu_1=\bar \mu,\qquad 
\mu_2=\alpha(\lambda_1,\lambda_2)\mu,\qquad
\bar\mu_2=\alpha(\bar\lambda_1,\bar\lambda_2)\bar\mu.
\eeq
We now use this parametrization and substitute (\ref{eq.Phiijvev}) and (\ref{eq.chiralF}) into 
(\ref{eq.simple}). We find that the $\alpha$ factors cancel and we get
\beq
\beta\partial_i\bar\partial_j\beta=\partial_i\beta\bar\partial_j\beta,
\eeq
where
\beq
\beta=\mu\bar\mu=\frac{m_1^2}{4}.
\eeq
This can be rewritten as
\beq
\partial_i\bar\partial_j\ln\beta=\frac{\partial_i\bar\partial_j\beta}{\beta}
-\frac{\partial_i\beta\bar\partial_j\beta}{\beta^2}=0,
\eeq
which means that $\ln\beta$ must be the sum of two chiral terms, 
\beq
\ln\beta=b(\lambda_1,\lambda_2)+b(\bar\lambda_1,\bar\lambda_2) , 
\eeq
and $\beta=\mu\bar\mu$ is chirally factorized. Thus we must have complete factorization:
\beq
\label{eq.comfact}
\mu=\mu(\lambda_1,\lambda_2),\qquad\quad
\bar\mu=\mu(\bar\lambda_1,\bar\lambda_2).
\eeq

\subsection{Mass-coupling relation}

Similarly, with the help of the generalized $\Theta$ sum rule, 
a Ward identity for the perturbing operators $\Phi_{ij}$ gives
a differential equation, from which the exact mass-coupling relation
is derived.

We will make use of the short distance expansion
\begin{equation}
J^-(z)\phi_i(w)=-\frac{M_{ij}\phi_j(w)}{(z-w)^2}+{\rm O}\left(\frac{1}
{z-w}\right), 
\label{eq.Mij}
\end{equation} 
where
\begin{equation}
M_{11}=1,\quad M_{12}=M_{21}=\frac{\eta}{2},\quad M_{22}=0,\qquad\eta=
\frac{\lambda_1}{\lambda_2}.
\end{equation} 
It is easy to see that
\begin{equation} 
\lambda_i M_{ij}=\frac{3}{2\pi} v_j.
\end{equation} 
For later use we calculate
\begin{equation}
Q_i=\pi v_k M_{ki}+v_k\partial_k v_i=\pi\left(2-\frac{\eta^2}{3}\right)v_i+
\frac{\pi^2\eta^2}{2}\lambda_i.
\end{equation} 
The generalized $\Theta$ sum rule corresponding to (\ref{eq.Mij}) is
\begin{equation}
\int {\rm d}^2 x\langle J^+(x)\Phi_{ij}(0)\rangle_c=\pi M_{ik}
\langle \Phi_{kj}\rangle,\qquad\quad J^+=v_i\Psi_i.
\end{equation} 

Let us consider the Ward identity
\begin{equation}
\partial_i\langle\Phi_{kj}\rangle=\int {\rm d}^2x\langle \Psi_i(x)
\Phi_{kj}(0)\rangle_c.
\end{equation} 
If we multiply this identity with $\lambda_i$, we get the original $\Theta$
sum rule. To obtain something new, we have to multiply with $v_i$. We then get
\begin{equation}
v_i\partial_i(\langle\Phi_{kj}\rangle)=
\int {\rm d}^2 x\langle J^+(x)\Phi_{kj}(0)\rangle_c=\pi M_{ki}
\langle \Phi_{ij}\rangle.
\end{equation} 
 Here we substitute into  $\Phi_{ij}$ the known relation between 
the  UV and IR fields (\ref{eq.PhiNX}) and (\ref{eq.Nabij}).
Factoring out  $\bar\partial_j\ln m_a$ depending only on $\bar\lambda_j$ from both sides, 
\begin{equation}
v_i\partial_i\left((\partial_k\ln m_b)
\langle X^{+-}_{(a)(b)}\rangle\right)=
\pi M_{ki}(\partial_i\ln m_b)
\langle X^{+-}_{(a)(b)}\rangle.
\end{equation} 
Now we use 
the VEVs of $X^{+-}_{(a)(b)}$ in (\ref{eq.vev1}), (\ref{eq.vev2}) and find
\begin{equation}
\tilde{v}_i\partial_i(\partial_j\ln m_x)+2\tilde{k}_x(\partial_j\ln m_x) 
= M_{ji}(\partial_i\ln m_x), 
\end{equation} 
where there is no summation over the index $_x$, 
which can be either $1$ or $2$,
and we have introduced
\begin{equation}
\tilde{k}_x=\tilde{v}_i\partial_i\ln m_x \, , \qquad 
 v_i=\pi \tilde{v}_i .
\end{equation} 
Multiplying this with $\lambda_j$ gives nothing new (the identity $-\tilde{k}_x+(5/2)\tilde{k}_x=
(3/2)\tilde{k}_x$). Multiplying with $\tilde{v}_j$ gives
\begin{equation}
\label{eq.dk0}
\tilde{v}_i\partial_i \tilde{k}_x+2\tilde{k}_x^2=(\tilde{v}_j\partial_j \tilde{v}_i
+ \tilde{v}_j M_{ji})\partial_i\ln m_x= \left(2-\frac{\eta^2}{3}\right)\tilde{k}_x
+\frac{5}{8}\eta^2.
\end{equation} 
Let us introduce the differential operator $ D=\tilde v_i\partial_i $ 
which acts on functions of $\eta$ as
\begin{equation}
Df(\eta)\equiv\tilde v_i\partial_i f(\eta)=\left(\eta-\frac{\eta^3}{3}\right)f^\prime(\eta).
\end{equation} 
Using dimensional analysis we can parametrize the chiral masses 
(in the fundamental domain defined in (\ref{eq.fndm}) below) as
\begin{equation}
\label{eq.muq}
2\mu_x=\lambda_1^{5/2}q_x(\eta).
\end{equation} 
In this parametrization,
\begin{equation}
2\tilde k_x = D \ln \mu_x
=\frac{5}{2}+\frac{Dq_x}{q_x} \, , 
\end{equation} 
and the differential equation (\ref{eq.dk0}) translates into
\begin{equation}
D^2q_x+\left(3+\frac{\eta^2}{3}\right)Dq_x
+\frac{5}{4}\left(1-\frac{\eta^2}{3}\right)q_x=0,
\end{equation} 
which can be simplified to
\begin{equation}
\label{eq.diff}
\eta^2\left(1-\frac{\eta^2}{3}\right)q^{\prime\prime}_x
+\eta\left(4-\frac{2\eta^2}{3}\right)q^\prime_x+\frac{5}{4}q_x=0.
\end{equation} 
This is a differential equation of hypergeometric type. Its solutions can be
expressed in terms of hypergeometric functions.

\subsection{Solution of the differential equation}
\label{subsec.solDE}

The differential equation (\ref{eq.diff}) has three regular singular
points at $\eta=0,\pm\sqrt{3}$. The exponents at the critical points are
$-1/2,-5/2$ (at $0$) and $0,2$ (at $\pm\sqrt{3}$). One solution of 
(\ref{eq.diff}) is
\beq
\left(\frac{\eta}{\sqrt{3}+\eta}\right)^{-1/2}{}_2F_1\left(-\frac{1}{2},
\frac{3}{2};3;\frac{2\eta}{\sqrt{3}+\eta}\right).
\label{eq.regsol}
\eeq
This behaves like $\eta^{-1/2}$ for $\eta\to0$. The other, more singular
solution goes like $\eta^{-5/2}$ for $\eta\to0$. Naively it would be given
by
\beq
\left(\frac{\eta}{\sqrt{3}+\eta}\right)^{-5/2} {}_2F_1\left(-\frac{5}{2},
-\frac{1}{2};-1;\frac{2\eta}{\sqrt{3}+\eta}\right),
\eeq
but this is ill-defined and the other solution must be given differently.
Luckily, the solution we need can also be expressed with the hypergeometric
function appearing in the first solution, at a different argument. We will
write, for short, 
\beq
\label{eq.Fz}
F(z)={}_2F_1\left(-\frac{1}{2},\frac{3}{2};3;z\right).
\eeq
For later use we note that
\beq
F(1)=\frac{32}{15\pi},\qquad\quad
F(1/2)=\frac{32\sqrt{\pi}}{5}\frac{1}{\Gamma^2(1/4)}.
\eeq

We now look for solutions which satisfy 
boundary conditions coming 
from the special cases discussed in Section \ref{subsec.special},
and are symmetric under the reflections and $\ZZ_3$ rotations
($S_3$ Weyl symmetry)
shown in Section \ref{subsec.cpt}. The latter conditions
are
\beq
\mu_a(\lambda_1,\lambda_2)=\mu_a(-\lambda_1,\lambda_2) ,
\eeq
and
\beq
\mu_a(\lambda_1,\lambda_2)=
\mu_a\left(\frac{\sqrt{3}}{2}\lambda_2-\frac{1}{2}\lambda_1,
-\frac{1}{2}\lambda_2-\frac{\sqrt{3}}{2}\lambda_1\right).
\eeq
The special case we need is the left-right 
symmetric point with couplings
$\lambda_1=\bar\lambda_1=0$, $\lambda_2=\bar\lambda_2=\lambda$. This model
is (up to identification of fields, as discussed in Section \ref{sec.HSGpCFT})
\beq
{\rm Ising\ (unperturbed)\ }\otimes\ {\rm perturbed\  tricritical\  Ising},
\eeq
where the perturbation is given in the tricritical Ising part by 
$ \nu \phi_{1,3}$ where $\nu=\lambda^2$. 
In this model one of the masses vanishes. By convention, we call this $m_1$.
The mass-coupling relation between $\lambda$ and the other, non-vanishing, mass
$m_2$ is known as in (\ref{eq.MCrsos}). Thus at this special point
\beq
m_1(0,\lambda\vert 0,\lambda)=0,\qquad\quad
m_2(0,\lambda\vert 0,\lambda)=K\lambda^{5/2},
\label{special1}
\eeq
where
\beq
  K = (\kappa_3^{\rm RSOS})^{-5/4} \comma
 \eeq
with $\kappa_3^{\rm RSOS}$ given in (\ref{eq.kappa3}).

The point $(0,-\lambda\vert 0,-\lambda)$ in coupling space
is obviously the same model, but here we have two possibilities. Either 
(case $a$) 
\beq
m_1(0,-\lambda\vert 0,-\lambda)=0,\qquad\quad
m_2(0,-\lambda\vert 0,-\lambda)=K\lambda^{5/2},
\label{special2a}
\eeq
or (case $b$)
\beq
m_2(0,-\lambda\vert 0,-\lambda)=0,\qquad\quad
m_1(0,-\lambda\vert 0,-\lambda)=K\lambda^{5/2}.
\label{special2b}
\eeq
For $\mu_1$, the solution satisfying the boundary condition (\ref{special1})
is based on (\ref{eq.regsol}):
\beq
\mu_1(\lambda_1,\lambda_2)=B\lambda_1^2 \lambda_2^{1/2}
(\sqrt{3}+\eta)^{1/2} F\left(\frac{2\eta}{\sqrt{3}+\eta}\right),
\label{eq.mu1}
\eeq
where $B$ is some constant. This solution is valid in the fundamental domain
\beq
\label{eq.fndm}
\lambda_2\geq\frac{\lambda_1}{\sqrt{3}}\geq0.
\eeq
Later we will also use the anti-fundamental domain defined by
\beq
\lambda_2\leq-\frac{\lambda_1}{\sqrt{3}}\leq0.
\eeq
Using the $\ZZ_3$ symmetry, we now  calculate
\beq
\mu_1(0,-\lambda)=\mu_1\left(\frac{\sqrt{3}}{2}\lambda,\frac{1}{2}\lambda\right)
=\frac{8}{5\pi}3^{1/4}B\lambda^{5/2}\not=0.
\eeq
This excludes case $a$ and we are left with
\beq
2\mu_1(0,-\lambda)=K\lambda^{5/2},
\eeq
which can be used to fix the constant $B$ as 
\beq
B=\frac{5\pi}{16}3^{-1/4}K.
\label{eq.B}
\eeq

Since we are left with case $b$, we can write down $\mu_2$ immediately using
the boundary condition and the differential equation. In the anti-fundamental
domain it takes the form
\beq
\mu_2(\lambda_1,\lambda_2)=\tilde B\lambda_1^2 (-\lambda_2)^{1/2}
(\sqrt{3}-\eta)^{1/2} F\left(\frac{-2\eta}{\sqrt{3}-\eta}\right).
\label{eq.mu2}
\eeq
This solution is based on the observation that $q^{\rm refl}_a(\eta)=q_a(-\eta)$
satisfies the same differential equation.
Using the $\ZZ_3$ symmetry, we can rotate this solution to the fundamental domain,
where it takes the form
\beq
\mu_2(\lambda_1,\lambda_2)=\frac{\tilde B}{4}(\sqrt{3}\lambda_2-\lambda_1)^2 
(\sqrt{3}\lambda_2+\lambda_1)^{1/2}
F\left(\frac{\sqrt{3}\lambda_2-\lambda_1}
{\sqrt{3}\lambda_2+\lambda_1}\right).
\label{eq.mu1fund}
\eeq
The boundary condition
\beq
2\mu_2(0,\lambda)=K\lambda^{5/2}
\eeq
tells us that $\tilde B=B$.
Incidentally, from (\ref{eq.mu1fund}) we can read off the other, more
singular solution of the differential equation:
\beq
q_2(\eta)=\frac{B}{2}\eta^{-5/2}(\sqrt{3}-\eta)^2(\sqrt{3}+\eta)^{1/2}
F\left(\frac{\sqrt{3}-\eta}{\sqrt{3}+\eta}\right).
\eeq

To summarize, the mass-coupling relation (in the fundamental region) is 
\beq
\label{eq.mulam}
\begin{split}
\mu_1(\lambda_1,\lambda_2)&=
B\lambda_1^2(\lambda_1+\sqrt{3}\lambda_2)^{1/2}F\left(
\frac{2\lambda_1}{\lambda_1+\sqrt{3}\lambda_2}\right),\\
\mu_2(\lambda_1,\lambda_2)&=
\frac{B}{4}(\sqrt{3}\lambda_2-\lambda_1)^2
(\lambda_1+\sqrt{3}\lambda_2)^{1/2}F\left(
\frac{\sqrt{3}\lambda_2-\lambda_1}{\lambda_1+\sqrt{3}\lambda_2}\right),
\end{split}
\eeq
where the constant $B$ is given by (\ref{eq.B}).
These expressions are extended outside the fundamental domain by the $S_3$ Weyl symmetry.
This is the main result in this paper.
Figure \ref{fig.mulam1} is a plot of $\mu_a(\lambda_i)$.
In Appendix \ref{app.e}, we summarize the symmetry of $\mu_a(\lambda_i)$ 
and their parametrization invariant under the symmetry.

\begin{figure}[t]
\begin{center}
\begin{tabular}{cc}
\resizebox{75mm}{!}{\includegraphics{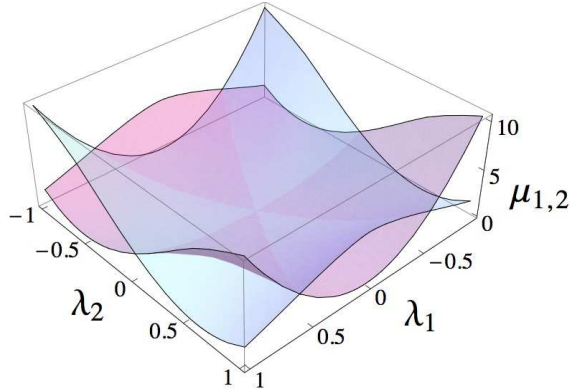}}
\vspace{-3mm}
\end{tabular}
\end{center}
\caption{Plot of $\mu_a(\lambda_i)$ in (\ref{eq.mulam}) extended to the entire 
$(\lambda_1,\lambda_2)$-plane. The red and the blue surface represent $\mu_1(\lambda_i)$
and $\mu_2(\lambda_i)$, respectively.
}
\label{fig.mulam1} 
\end{figure}

Having found the solution, we can now study the other special case. It is easy
to see that
\beq
\mu_1\left(\frac{1}{2}\lambda,\frac{\sqrt{3}}{2}\lambda\right)
=\mu_2\left(\frac{1}{2}\lambda,\frac{\sqrt{3}}{2}\lambda\right)
=\frac{B}{2\sqrt{2}}F(1/2)\lambda^{5/2}.
\eeq
The point $(\lambda/2,\sqrt{3}\lambda/2)$ in coupling space can be transformed to
$(\lambda,0)$ by a reflection followed by a $120$ degree rotation. Thus
\beq
2\mu_1(\lambda,0)=2\mu_2(\lambda,0)=\tilde K\lambda^{5/2},
\eeq
with
\beq
\label{eq.Ktilde}
\tilde K=\frac{B}{\sqrt{2}}F(1/2).
\eeq
We can calculate the ratio $\tilde K/K$ analytically. We  find
\beq
\frac{\tilde K}{K} 
=3^{-1/4}\frac{\pi\sqrt{2\pi}}{\Gamma^2(1/4)}.
\eeq
One can check that 
this is the same as found in \cite{Hatsuda:2011ke} representing this special case
as perturbation of the non-unitary minimal model ${\cal M}_{3,5}$.
The constant $\tilde{K}$ is related to $\hat\kappa$ in (\ref{eq.kappahat})
as $\tilde{K}^{-16/5} =  (\hat\kappa/\pi)^2  \gamma(9/5) \gamma^3(2/5)$.

\section{Numerical mass-coupling relation}
\label{sec.NumMCrel}

We have found the exact mass-coupling relation $\mu_a = \mu_a(\lambda_i)$  in (\ref{eq.mulam}).
In this section, we summarize our numerical investigations of  the TBA system and 
the mass-coupling relation, providing numerical checks of our analytic findings
so far. We then discuss the inverse mass-coupling relation $ \lambda_i = \lambda_i(\mu_a)$
with the help of numerics, which is necessary to express the pCFT results in terms of $\lambda_i$
by  the IR variables. We also make a comment on an earlier work on the mass-coupling relation.

\subsection{UV expansion coefficients of the ground state energy from TBA}
\label{sec.uvexp}

First, we present the outcome of our numerical investigations of the TBA equations.
By  solving the TBA equations (\ref{eq.tba1b}) and (\ref{eq.tba2b}) numerically and using (\ref{eq.fl2})
one can determine $F(L)$ at different 
values of 
$L$, and then extract from these results $c$ and the first few coefficients $F_n$ appearing in 
(\ref{eq.uvexp}). 

The pCFT formulation of the model implies $c=6/5$ and $F_1=0$, and the numerical values that 
can be obtained from the TBA agree with these exact values with high accuracy.

Equation (\ref{eq.fw4}) and the chiral factorization derived in Section \ref{subsec.compfactr} 
imply that $F_2$ and $F_3$ have the following factorization properties:
\bea
\label{eq.fact1}
F_2(\mu_1, \mu_2, \bar{\mu}_1,\bar{\mu}_2) 
& = & \phantom{-}G_2(\mu_1,\mu_2)G_2(\bar{\mu}_1,\bar{\mu}_2) \ ,\\
\label{eq.fact2}
F_3(\mu_1, \mu_2, \bar{\mu}_1,\bar{\mu}_2) & = & -G_3(\mu_1,\mu_2)G_3(\bar{\mu}_1,\bar{\mu}_2)\ ,
\eea
where $G_2$ and $G_3$ are real. These properties can also be confirmed numerically. 

Regarding $F_4$, $F_5$ and $F_7$, we found that $F_4/F_2^2$, $F_5/(F_2 F_3)$ and $F_7/(F_2^2 F_3)$ 
are constant (i.e.\ they do not depend on  $\mu_i$, $\bar{\mu}_i$), 
again in agreement with the pCFT results in Section \ref{subsec.cpt}. 
The numerical values of $F_4/F_2^2$, $F_5/(F_2 F_3)$ and $F_7/(F_2^2 F_3)$ are
\beq
\frac{F_4}{F_2^2}=B_4\approx 0.33913,\qquad \frac{F_5}{F_2 F_3}=B_5\approx -1.1295,\qquad
\frac{F_7}{F_2^2 F_3}=B_7\approx 1.685\ .
\eeq
In order to compare these values with (\ref{eq.const}) it would be necessary to calculate the latter
constants in pCFT.
These results together with (\ref{eq.fact1}), (\ref{eq.fact2}) imply that 
$F_4, F_5, F_7$ are also factorized,
\bea
\label{eq.fact4}
F_4(\mu_1, \mu_2, \bar{\mu}_1,\bar{\mu}_2) & = & \phantom{-}G_4(\mu_1,\mu_2)G_4(\bar{\mu}_1,\bar{\mu}_2) \ ,\\
\label{eq.fact5}
F_5(\mu_1, \mu_2, \bar{\mu}_1,\bar{\mu}_2) & = & \phantom{-}G_5(\mu_1,\mu_2)G_5(\bar{\mu}_1,\bar{\mu}_2) \ ,\\
\label{eq.fact7}
F_7(\mu_1, \mu_2, \bar{\mu}_1,\bar{\mu}_2) & = & -G_7(\mu_1,\mu_2)G_7(\bar{\mu}_1,\bar{\mu}_2)\ ,
\eea
and
$G_4/G_2^2$, $G_5/(G_2G_3)$, $G_7/(G_2^2G_3)$ are constant.

$F_6$ is not factorized, but instead it is found numerically to satisfy the more complicated relation
\bea
F_6(\mu_1, \mu_2, \bar{\mu}_1,\bar{\mu}_2) & = &
B_{622} F_2(\mu_1, \mu_2, \bar{\mu}_1,\bar{\mu}_2)^3 + 
B_{633} F_3(\mu_1, \mu_2, \bar{\mu}_1,\bar{\mu}_2)^2 \nonumber \\ 
&& + B_{623} G_2(\mu_1, \mu_2)^3 G_3(\bar{\mu}_1,\bar{\mu}_2)^2 \nonumber\\
&& + B_{632} G_3(\mu_1, \mu_2)^2 G_2(\bar{\mu}_1,\bar{\mu}_2)^3,
\label{eq.fact6}
\eea
where
\beq
B_{622}\approx -0.0745 ,\qquad B_{633}\approx 0.2221,\qquad B_{623}= B_{632}\approx 0.2704\ .
\eeq
Clearly, the structure of (\ref{eq.fact6}) is similar to that of (\ref{eq.f6b}). 
Taking into consideration the previous results, it can be seen that (\ref{eq.fact6}) follows from (\ref{eq.f6b}) if 
\beq
B_{622}=\frac{C_{622}}{C_2^3},\qquad B_{633}=\frac{C_{633}}{C_3^2},\qquad
B_{623}=-\frac{C_{623}}{C_3 C_2^{3/2}}\ .
\eeq

In addition to the symmetries listed below (\ref{eq.fl2}), we found numerically that 
\bea
\label{eq.sign1}
F_2(\mu_1,\mu_2,\bar{\mu}_1,\bar{\mu}_2) & = & \phantom{-}F_2(\mu_2,\mu_1,\bar{\mu}_1,\bar{\mu}_2) \ ,\\
\label{eq.sign2}
F_3(\mu_1,\mu_2,\bar{\mu}_1,\bar{\mu}_2) & = & -F_3(\mu_2,\mu_1,\bar{\mu}_1,\bar{\mu}_2)\ .
\eea
These are indeed derived on the  pCFT side in the next subsection.
Due to the Dynkin reflection symmetry the same transformation rules apply under 
$\bar{\mu}_1 \leftrightarrow \bar{\mu}_2$. For $G_2$ and $G_3$ these properties imply
\bea
\label{eq.sign2.4}
G_2(\mu_1,\mu_2) & = & \phantom{-}G_2(\mu_2,\mu_1) \ , \\
\label{eq.sign2.5}
G_3(\mu_1,\mu_2) & = & -G_3(\mu_2,\mu_1)\ .
\eea
From the relations between $F_2$, $F_3$ and $F_4$, $F_5$, $F_6$, $F_7$ described above 
it follows then that
\beq
\label{eq.sign3}
F_n(\mu_1,\mu_2,\bar{\mu}_1,\bar{\mu}_2)  = 
(-1)^n F_n(\mu_2,\mu_1,\bar{\mu}_1,\bar{\mu}_2),\qquad n=4,\dots, 7 \ ,
\eeq
\bea
\label{eq.sign6}
&& G_n(\mu_1,\mu_2)\, =\,  \phantom{-}G_n(\mu_2,\mu_1),\qquad n=2,4 \ , \\
\label{eq.sign7}
&& G_n(\mu_1,\mu_2)\, =\, -G_n(\mu_2,\mu_1),\qquad n=3,5,7\ .
\eea

Finally, 
in the case when $\mu_i=\bar{\mu}_i$ we found that
$F_3^2/F_2^3$ grows monotonically from $0$ to $\sim\frac{C_3^2}{C_2^3}=5.26554\dots$
as $m_1/m_2$ goes from $1$ to $0$. This result is consistent with (\ref{eq.ineq}).

\subsection{Numerical mass-coupling relation}
 
Let us move on to the numerical investigation of the mass-coupling relation.
Given the numerical coefficients $F_n$ as in the previous subsection,
the equations (\ref{eq.p2}) and (\ref{eq.fw4}) determine the couplings  $\lambda_i, \bar\lambda_i $.
By the factorization 
(\ref{eq.comfact}), they are functions of $\mu_a$ and $ \bar\mu_a$, respectively.
Setting $\mu_a = \bar\mu_a$ for simplicity,
one obtains twelve sets of the solutions $(\lambda_1,\lambda_2)$, which are indeed real.
Due to the $S_3$ Weyl symmetry, 
without loss of generality  $(\lambda_1,\lambda_2)$ are set 
to be in the fundamental domain $\lambda_2\geq\frac{\lambda_1}{\sqrt{3}}\geq0$,
or the fundamental Weyl chamber of the $su(3)$ weight space.
The resultant two sets in this domain are regarded as a pair related by 
the Dynkin reflection $ (\mu_1, \bar\mu_1) \leftrightarrow (\mu_2, \bar\mu_2)$. 

\begin{figure}[t]
\begin{center}
\begin{tabular}{cc}
\resizebox{64mm}{!}{\includegraphics{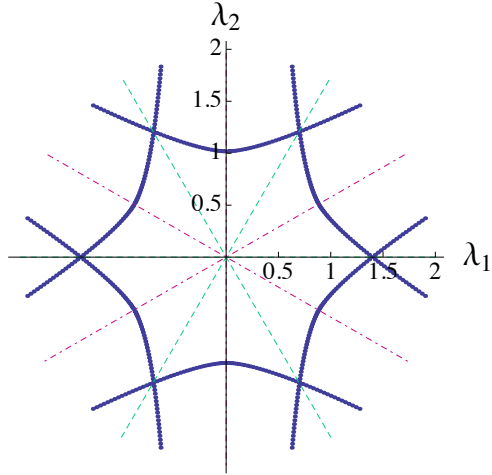}}
\vspace{-3mm}
\end{tabular}
\end{center}
\caption{Plot of numerical data of $\lambda_i$ from the TBA equations as function of  $\mu_1$  when 
 $(\mu_2)^{2/5} = 2$ is kept fixed. The dot-dashed lines represent the
$\mu_1 =0$ or $\mu_2 =0$ cases, while the dotted lines the $\mu_1 = \mu_2$ cases.
}
\label{fig.lambda} 
\end{figure}

Figure \ref{fig.lambda} is a plot of $\lambda_i$ obtained in this way, where 
$(\mu_2)^{2/5} = 2$ and $\mu_1$ is varied.
The points represent $\lambda_i(\mu_a)$ from the TBA equations.
The dot-dashed lines represent  $ \lambda_2 = \tan((2k+1)\pi/6) \lambda_1$ $(k=0,1,2)$,
whose intersections with the trajectories of the points correspond to the single-mass cases
$\mu_1 =0$ or $\mu_2 =0$.
The dotted lines represents $ \lambda_2 = \tan(k\pi/3) \lambda_1$ $(k=0,1,2)$, whose 
intersections at the cusps correspond to the equal-mass cases $\mu_1 = \mu_2$.
The twelve sets of the solutions form 
the twelve branches starting from the single-mass points 
(mid points of each edge of the hexagon).

\begin{figure}[t]
\begin{center}
\begin{tabular}{cc}
\hspace{-3mm}
\resizebox{70mm}{!}{\includegraphics{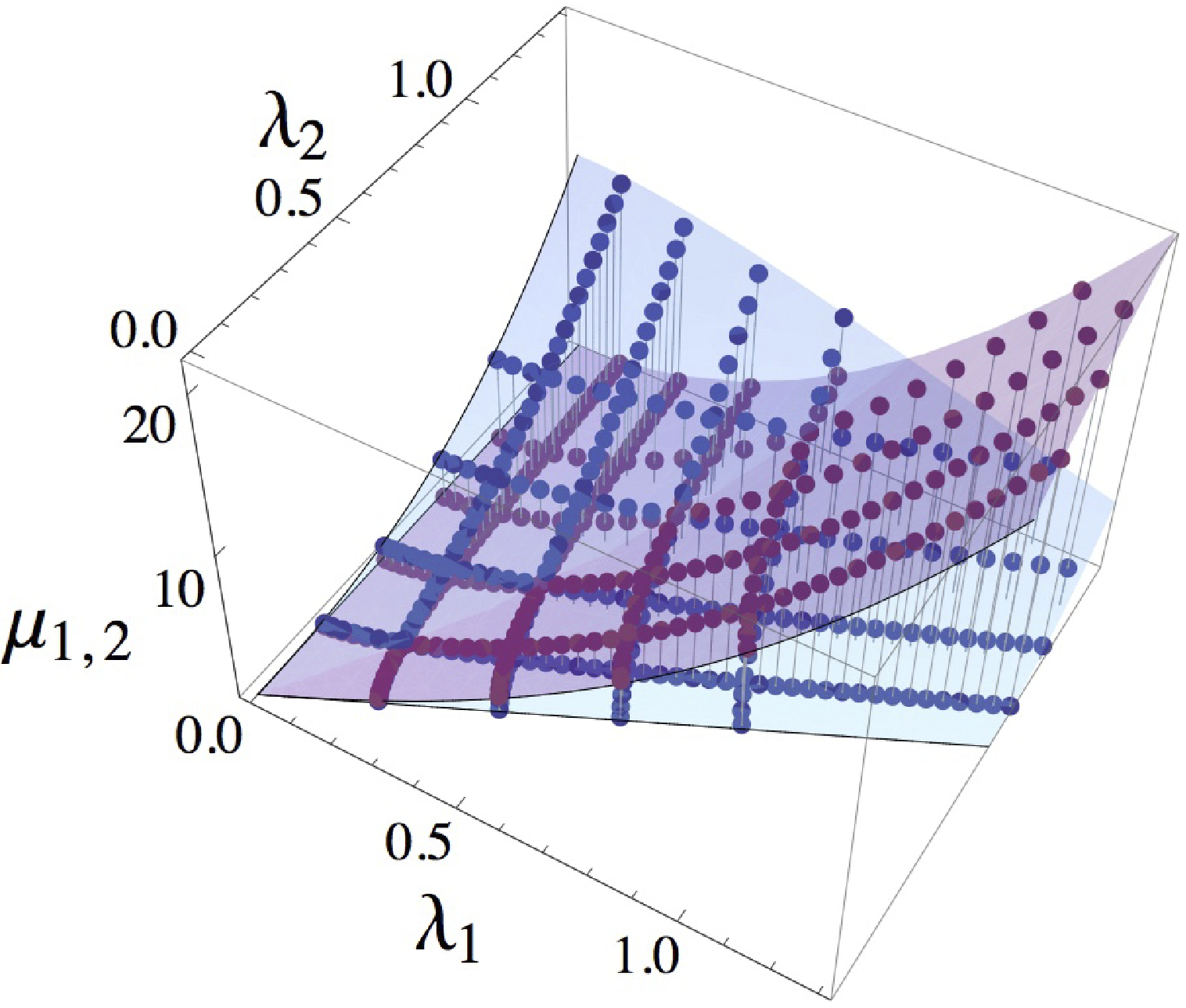}}
\hspace{8mm}
\vspace{1mm}
&
\resizebox{51mm}{!}{\includegraphics{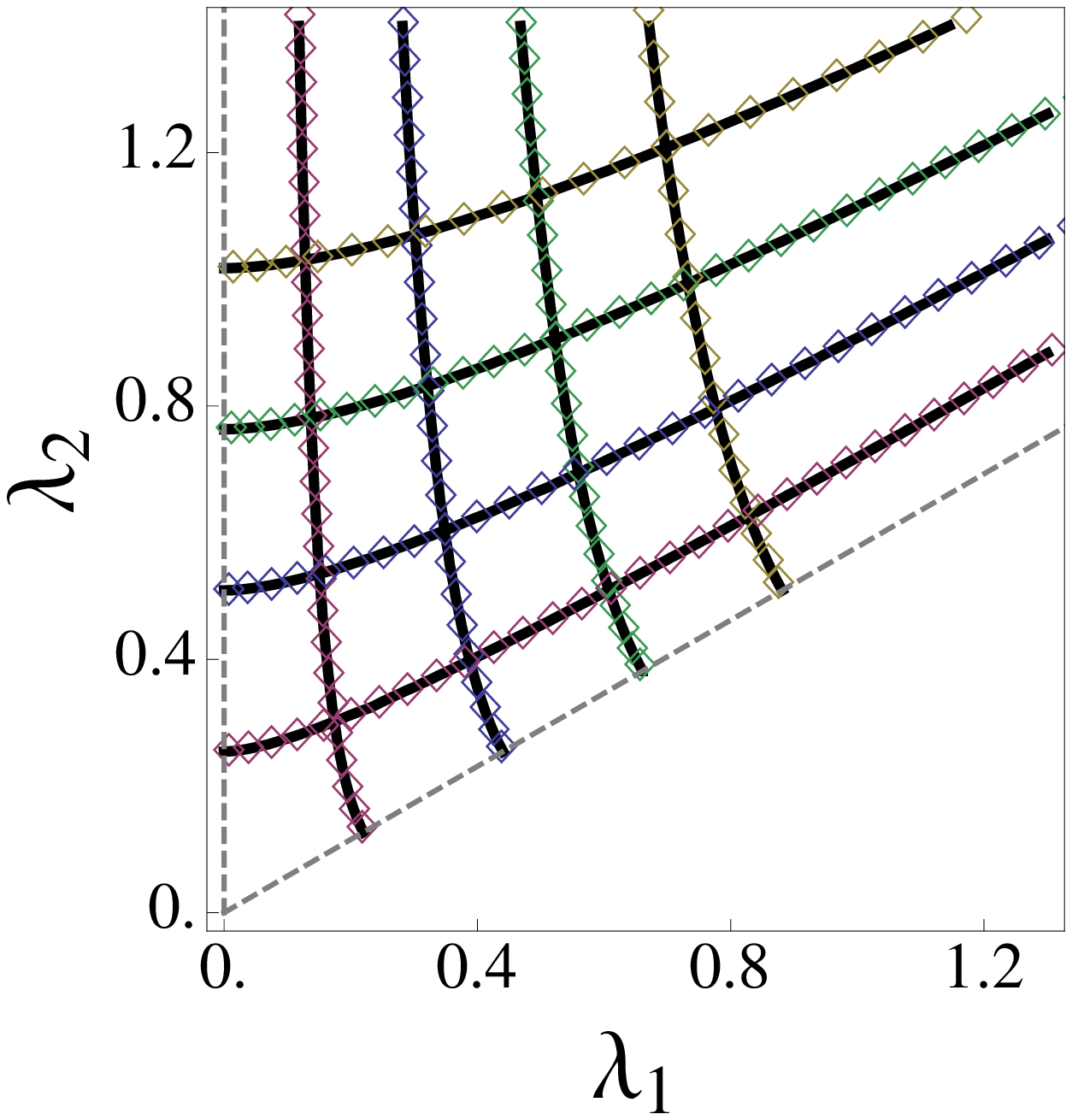}}
\hspace{0mm}
\\
(a) & (b)
\end{tabular}

\end{center}
\vspace{-3mm}
\caption{Comparison of the analytical and the numerical results for mass coupling relations. 
The red and blue  surfaces represent the analytical formula $\mu_a (\lambda_i)$ $(a=1,2)$ in (\ref{eq.mulam}),
respectively,  while red and blue points represent the numerical data.  On the right points are 
projected  to the $(\lambda_1, \lambda_2)$-plane.
}
\label{fig.mulam23}
\end{figure}

These  are compared with the analytic ones.
Figure \ref{fig.mulam23}  (a) is a plot of $(\mu_1, \mu_2)$ versus $(\lambda_1, \lambda_2)$
in the fundamental domain.
The red and blue  surfaces represent $\mu_a (\lambda_i)$ $(a=1,2)$ in (\ref{eq.mulam}),
respectively,  
whereas the red and blue points represent the numerical data 
$\lambda_i(\mu_a)$  for given $\mu_a (=\bar\mu_a)$.
Each horizontal sequence from the bottom 
to the top corresponds to $(\mu_2)^{2/5} = 1/2, 1, 3/2, 2$, with $\mu_1 $ varied,
while each vertical sequence from the left 
to the right corresponds to $(\mu_1)^{2/5} = 1/2, 1, 3/2, 2$, with $\mu_2 $ varied.
In Figure \ref{fig.mulam23} (b), the diamonds ($\diamond$) represent 
the projections of the points in (a) to the $(\lambda_1, \lambda_2)$-plane. 
The horizontal solid lines are the contours  of  
$  \bigl( \mu_2(\lambda_i) \bigr)^{2/5} = 1/2, 1, 3/2, 2$ from (\ref{eq.mulam}), while
the vertical solid lines are the contours  of  
$  \bigl( \mu_1(\lambda_i) \bigr)^{2/5} = 1/2, 1, 3/2, 2$.
We find good agreement between the analytic results and the numerical ones.

\subsection{Inverse relation}

The analytic mass-coupling relation (\ref{eq.mulam}) expresses the mass parameters
$\mu_a$ as functions of the couplings $\lambda_i$, whereas what one obtains numerically 
from the TBA equations is $\lambda_i$ as functions of $\mu_a$, i.e., the inverse relation
of (\ref{eq.mulam}). 
On dimensional grounds, a useful parametrization  in the fundamental domain is
\bea
 \label{eq.mulaminv}
  (\lambda_1,\lambda_2) =  \mu_1^{2/5} \rho_1(\xi) \cdot w_1 
  + \mu_2^{2/5} \rho_2(\xi) \cdot w_2 \comma 
 \eea
(and similarly for $\bar\lambda_i(\bar\mu_a)$), 
where $\rho_a \geq 0$, $w_i$ are the fundamental weighs of $su(3)$; $w_1=(\sqrt{3},1)/\sqrt{6}$, 
$w_2 = (0,2)/\sqrt{6}$; and 
\beq
\xi=\frac{\mu_1}{\mu_2}\ .\qquad 
\eeq
This parametrization generalizes, up to the power of $\mu_a$, 
a classical one in \cite{Dorey:2004qc}. 
 (\ref{eq.mulam}) implies that $\xi $ is a function of $\eta = \lambda_1/\lambda_2$.
 In Appendix \ref{app.f}, we show that $\xi(\eta)$ is a monotonically increasing 
 function in the fundamental domain,
 and its inverse $\eta(\xi)$ is well-defined.
 From the symmetry of the analytic relation
 under the chiral Dynkin transformation, $\mu_1 \leftrightarrow \mu_2$ with $\bar\mu_a$ fixed, 
 which is shown in Appendix \ref{app.e},
 it follows that 
\begin{equation}
\label{eq.rhoxi}
\rho_2(1/\xi)=\rho_1(\xi).
\end{equation}
The inverse mass-coupling relations thus can be expressed in terms of a single
function of one variable. 
Substituting (\ref{eq.mulaminv}) into (\ref{eq.p2}), 
one also finds $(p_2,p_3)\to (p_2,-p_3)$ 
under the chiral Dynkin transformation,
proving the relations (\ref{eq.sign1}) and (\ref{eq.sign2}) on the  pCFT side. 
If $\rho_a$ were unity,  (\ref{eq.mulaminv}) would give  
$\mu_1 \sim \lambda_1^{5/2} $, $\mu_2 \sim (\sqrt{3} \lambda_2 -\lambda_1)^{5/2}$.
The relations (\ref{eq.mulam}) are generalizing these. 
From (\ref{eq.mulaminv})  and (\ref{eq.mulam}), one finds that 
$\rho_a$ take a simple form $x_a^{-1/2} F[x_a]$ with $x_a(\lambda_j)$ being simple functions
of $\lambda_j$.

The differential equations for $\lambda_i$ or $\rho_a$ are derived by inverting the Jacobian
matrix $\del \mu_a/\del \lambda_i$, giving $\del \lambda_i/\del \mu_a$. 
It is, however, difficult to solve them generally. Instead, let us first consider the asymptotic
forms for 
 $\mu_1 \approx 0$ and $\mu_2 \approx 0$, corresponding to $\lambda_1 \approx 0$ 
 and  $ \sqrt{3} \lambda_2 \approx \lambda_1$, respectively.  
 From (\ref{eq.mulam}) it follows  that $F(1) \xi \approx (2\eta/\sqrt{3})^2$, 
 and hence
\beq
 \rho_1(\xi) \approx c_1 \xi^{1/10} + \calO(\xi^{3/5}) \comma \qquad 
 \rho_2(\xi) \approx c_2 (1- c_3 \xi^{1/2}) + o(\xi)  \comma 
\eeq
for  $ \xi \ll 1$.
Here, the constants $c_a$ are
\bea
 &&  c_1 = 2^{3/10} B^{-2/5} \bigl[ F(1) \bigr]^{1/10} \comma \qquad
    c_2 = c_1 \cdot  \bigl[ F(1)  \bigr]^{-1/2} \comma \nn \\
 &&    c_3 = \frac{2}{5}   \bigl[ F(1)  \bigr]^{-1/2} 
    \left( \frac{1}{2} F(1) - F'(1) \right)
   \period
\eea
Similarly, one has    
$F(1) \xi^{-1} \approx  \bigl((\sqrt{3} -\eta)/2 \eta \bigr)^2 $
 and  
\beq
   \rho_1(\xi) \approx  c_2 (1- c_3 \xi^{-1/ 2})  + o(\xi^{-1}) \comma \qquad  
 \rho_2(\xi) \approx  c_1 \xi^{-1/10} + \calO(\xi^{-3/5}) \comma 
\eeq
for $\xi \gg 1$.\footnote{$F(x)$ has a branch point at $x=1$, but $F'(x)$ exists  for $x \leq 1$ and Taylor's
theorem with Peano's form of the remainder can be applied.
The asymptotic behaviors are well approximated by functions of 
the form $\xi^{\pm1/10} \sum c_k \xi^{\pm k/2}$ or $\sum c_k \xi^{\pm k/2}$.}
From these,  the special values of $\rho_a$  are  read off,
\bea
\label{eq.sprho}
  \rho_1(0) & = & \rho_2(\infty) = 0 \comma \nn \\
  \rho_1(\infty) & = & \rho_2(0) = c_2 \approx 0.62317 \comma \\
  \rho_1(1) & = & \rho_2(1) = 2^{1/10} \bigl[ B F(1/2)\bigr]^{-2/5} \approx 0.49291 \period \nn
\eea
For reference, we have added the values at $\xi =1$, corresponding to $\lambda_2 = \sqrt{3} \lambda_1$.

\begin{figure}[t]
\begin{center}
\begin{tabular}{cc}
\resizebox{73mm}{!}{\includegraphics{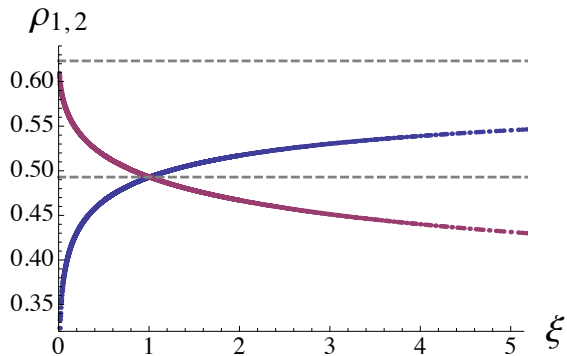}}
\vspace{-3mm}
\end{tabular}
\end{center}
\caption{ Plot for numerically inverting the mass-coupling relation. Blue points represent $\rho_1(\xi)$, whereas
the red points $\rho_2(\xi)$.
}
\label{fig.rho} 
\end{figure}

Generally, one can invert the relations (\ref{eq.mulam}) numerically.
It is confirmed 
 that the relation (\ref{eq.rhoxi}) indeed holds.
Figure \ref{fig.rho} is a plot of $\rho_a(\xi)$ obtained in this way.
The blue points in the increasing sequence represent $\rho_1(\xi)$, whereas
the red points in the decreasing sequence represent $\rho_2(\xi)$.
The dashed lines indicate the special values in (\ref{eq.sprho}).

\subsection{Comment on earlier work}

Finally, we comment on an earlier work \cite{Hatsuda:2011ke}, where
the mass-coupling relation of the $su(3)_2/u(1)^2$ HSG model
was studied in order to evaluate the strong-coupling amplitudes of ${\cal N }=4$ SYM.
Assuming that $\lambda_i$ are polynomials of $\mu_a^{2/5}$, the couplings
were parametrized as $\lambda_i = \sum_{a} \mu_a^{2/5} \hat\lambda_{ai} $.
The constants $\hat\lambda_{ai}$ were determined by matching the perturbative expression 
of $F_2$ in (\ref{eq.fw4}) and those in the perturbed minimal models corresponding to
the single-mass and equal-mass cases in Section \ref{subsec.special}.
The results were used for analytic expansions of the ground state energy and the Y-functions 
around the UV limit. It was observed that they appeared to be consistent 
with numerical data from the TBA equations within the numerical precision.

\begin{figure}[t]
\begin{center}
\begin{tabular}{cc}
\resizebox{75mm}{!}{\includegraphics{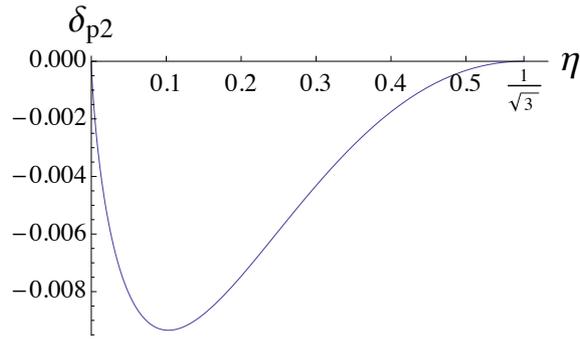}}
\vspace{-3mm}
\end{tabular}
\end{center}
\caption{Plot of the relative deviation of the exact expression and the one which was obtained  
assuming that $\lambda_i$ are polynomials of $\mu_a^{2/5}$.}
\label{fig.delta} 
\end{figure}

For the amplitudes, only the result of $F_2$ was used.
The chiral factor $p_2$ of $F_2$ in (\ref{eq.fw4}) reads there
\beq
  p_2^{HISS} = \frac{2}{3} \mu_2^{4/5} [ r_1 (1+ \xi^{4/5}) + r_2 \xi^{2/5}] \comma
\eeq
where 
\beq
   r_1= \rho_2^2(0) \comma \qquad r_2 = 3 \rho^2_2(1) - 2 r_1 \period
\eeq
With these constants, $p_2^{HISS}$ indeed matches the expression from (\ref{eq.mulaminv}),
\beq
  p_2 =  \frac{2}{3} \mu_2^{4/5} [ \rho_2^2 + \rho_1 \rho_2 \xi^{2/5} + \rho_1^2 \xi^{4/5}] \comma
\eeq 
at $\xi = 0,1$. Figure \ref{fig.delta} is a plot of the relative deviation of the two expressions, 
\beq
   \delta_{p2} = \frac{p_2^{HISS}}{p_2} -1 \period
\eeq
For simplicity, $\delta_{p2}$ is shown as a function of $\eta$ in the range 
$ 0 \leq \eta = \lambda_1/\lambda_2 \leq 1/\sqrt{3}$,
corresponding to $ 0 \leq \xi \leq 1$.
The case with $ \xi > 1$ is covered by the Dynkin symmetry.
One finds that the deviation is  less than 1 per cent.
It is still an open problem why a simple assumption in \cite{Hatsuda:2011ke} works so well effectively.
Other part of the analyses in \cite{Hatsuda:2011ke} does not depend on the exact form of $F_2$
and hence need not be corrected. Similar remarks may apply to the analyses in 
\cite{Hatsuda:2012pb} for the $su(3)_4/u(1)^2$ HSG model.

\section{Vacuum expectation values from the mass-coupling relation}
\label{sec.VEV}

Given the mass-coupling relation, one can obtain
the vacuum expectation values of the perturbing operators,
which are the derivatives of the partition function 
with respect to the couplings. 
Indeed, in the course of  the derivation of the analytic mass-coupling relation, 
a number of formulas have been found from the UV as well as the IR side:
 (\ref{eq.thetavev}), (\ref{eq.firstder}),  (\ref{eq.ABCvev}),
(\ref{eq.Dvev}) and (\ref{eq.simple}).

\begin{figure}[t]
\begin{center}
\begin{tabular}{cc}
\hspace{-3mm}
\resizebox{69mm}{!}{\includegraphics{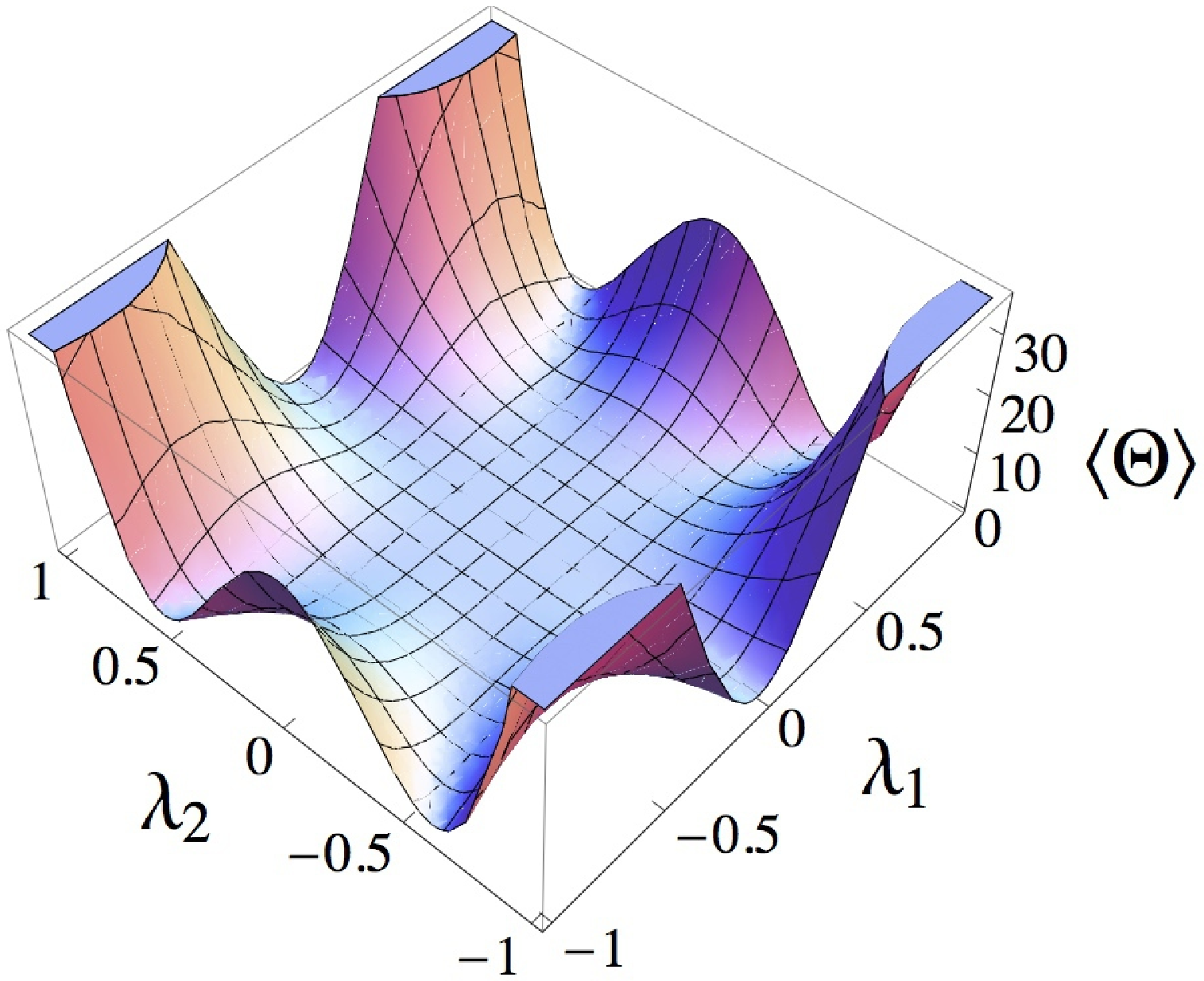}}
\hspace{8mm}
\vspace{1mm}
&
\resizebox{66mm}{!}{\includegraphics{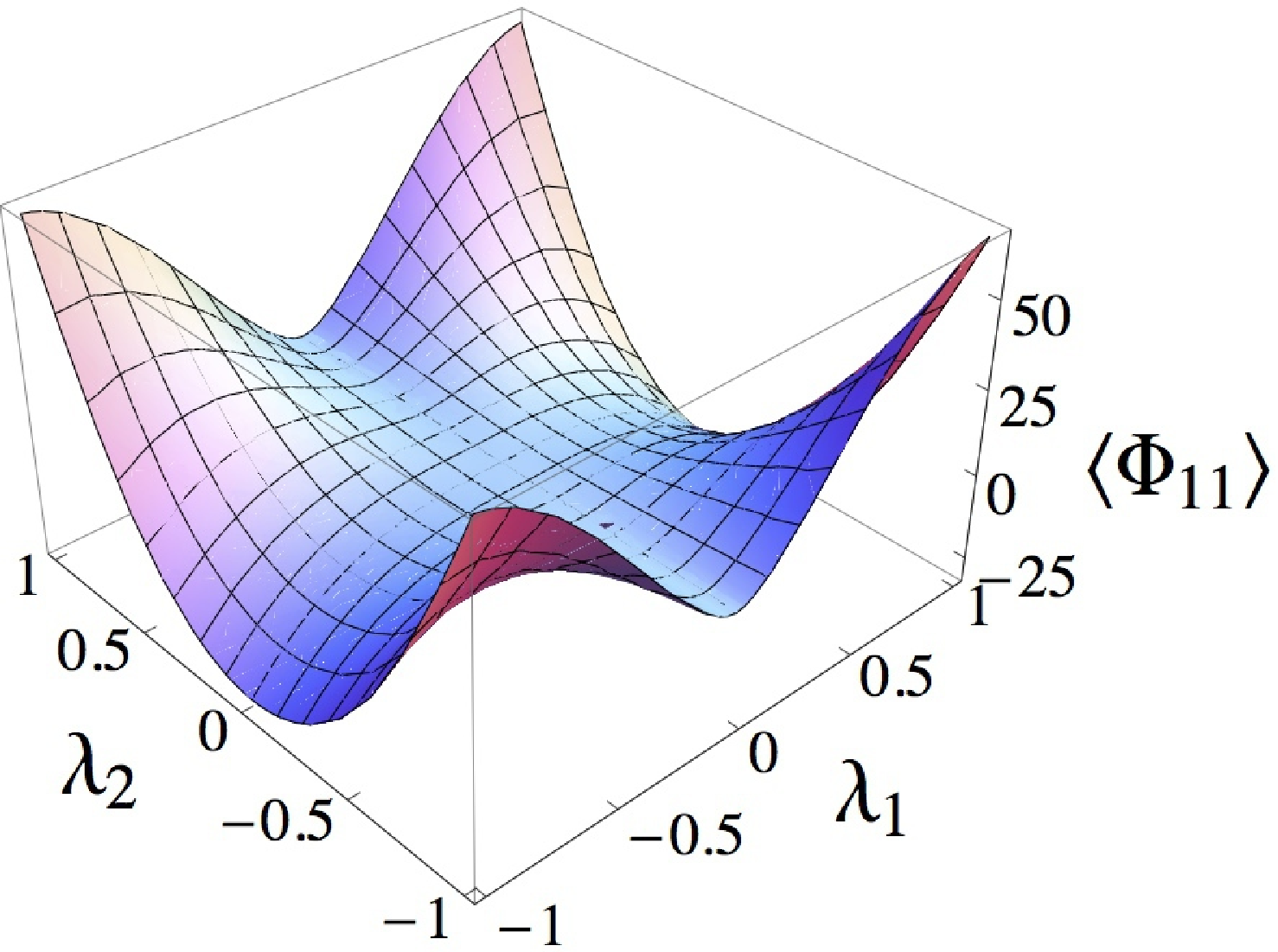}}
\hspace{0mm}
\\
(a) & (b)
\end{tabular}

\end{center}
\vspace{-3mm}
\caption{Plots of (a) $\langle \Theta \rangle$  and (b) $\langle \Phi_{11} \rangle$
as functions of  $\lambda_j = \bar\lambda_j$. 
}
\label{fig.vevlam}
\end{figure}

 To be concrete,  $\langle \Phi_{ij} \rangle $ 
are for example given in terms of the couplings 
$\lambda_i,\bar\lambda_j$ by (\ref{eq.simple}), while those in terms of $\mu_a, \bar\mu_b$
are obtained through the inverse relation (\ref{eq.mulaminv}).
On the other hand, $\langle \Theta \rangle $ is simply given by the mass parameters 
as in (\ref{eq.thetavev}), which is expressed by $\lambda_i,\bar\lambda_j$
through the mass-coupling relation, e.g.,  (\ref{eq.mulam}) in the fundamental domain.
In Figure \ref{fig.vevlam} 
(a) and (b), we show plots of the vacuum expectation values as functions of the couplings, 
for examples of $\langle \Theta \rangle $ and $\langle \Phi_{11} \rangle $.
For simplicity, we have set $\lambda_j = \bar\lambda_j$.
 
\section{Conclusions}
\label{sec.concl}

In this paper, we studied the mass-coupling relation of multi-scale 
quantum integrable models, focusing on the $su(3)_2/u(1)^2$ HSG model
as their simplest example. Our basic strategy is to compare the 
conservation laws and the Ward identities of the integrable model both from the UV 
and IR points of view,
which provides a novel method to analyze integrable models.

For this purpose, we first identified the relevant conserved currents 
on the UV side, and the dimension 3/5 operators on the IR side, 
which are the counterpart of the UV perturbing operators and characterized 
by their form factors. 
The representation of the coset in terms of the projected product of the minimal models
provided  an efficient calculational basis. 
 It is notable that the products of minimal models allow multi-parameter integrable perturbations.
Using the formulas for the response of the masses and S-matrix under 
the variation of the couplings, the perturbing operators $\Psi_i $ were expressed by
the IR operators. This enabled us to express the conserved currents in terms of the IR operators.
Comparing the conservation laws on the UV and the IR sides, the perturbing operators $\Phi_{ij}$
were also expressed by the IR operators. From the generalized $\Theta$ sum rule and the free energy 
Ward identity, the factorization of the mass-coupling relation (\ref{eq.comfact}) was shown.

The Ward identity for $\Phi_{ij}$ gave a differential equation of their one-point function.
Together with the IR expression of $\Phi_{ij}$, it was translated into a differential equation 
for the mass-coupling relation, which led to our main result (\ref{eq.mulam}).
In the course of the derivation, we also obtained the vacuum expectation values of
the perturbing operators.  
The resultant mass-coupling relation reproduced the known exact results in the single-mass
cases, and agreed with the data obtained by solving the TBA equations numerically.
Via the gauge-string duality, the relation provides the missing link to develop 
an analytic expansion of ten-particle strong-coupling scattering amplitudes
of $\mathcal{N}=4$ SYM around the $\bbZ_{10}$-symmetric (regular-polygonal)
kinematic point.

Though we concentrated on the $su(3)_2/u(1)^2$ HSG model,
our discussion in this paper is conceptually more general 
and can be applied to other multi-scale integrable models. 
Once a set of relevant form factors are given, the analysis of the mass-coupling relation 
would be straightforward. Our derivation also implies that one can obtain the differential equation 
for the one-point functions of the perturbing operators only through the UV
conserved currents. Recalling the importance of differential equations 
in determining the correlations functions at the critical point, 
it would be an interesting future problem 
how powerful this Ward identity/differential equation is in determining the non-perturbative
off-critical one-point functions.

\acknowledgments
We would like to thank J. Luis Miramontes for useful conversations 
and L\'aszl\'o Feh\'er for a discussion on the Weyl reflection group.
This work was supported by Japan-Hungary Research Cooperative Program.
Z.~B., J.~B. and  G.~Zs.~T. were supported by a Lend\"ulet Grant and by OTKA 
K116505,
whereas K.~I. and Y.~S. were supported by JSPS Grant-in-Aid for Scientific Research,
15K05043 and 24540248 from Japan Society for the Promotion of Science (JSPS).

\appendix

\section{Conventions}
\label{app.convention}
 
In this appendix, we summarize our conventions.

\subsection{Space-time coordinates}

We use the Minkowski space coordinates
\begin{equation}
x\sim(x^0,x^1)=(t,x),\qquad\quad
x^\pm=t\pm x ,
\end{equation} 
and for any 2-vector $W^\mu$ we define
\begin{equation}
W^\pm=W^0\pm W^1 , \qquad\quad W_\pm=\frac{1}{2}W^\mp.
\end{equation} 
The scalar product is
\begin{equation}
V\cdot W=V^+W_++V^-W_-=\frac{1}{2}(V^+W^-+V^-W^+), \qquad\quad
x^2=x^+x^- .
\end{equation} 
The derivatives are given as
\begin{equation}
\partial_0=\frac{\partial}{\partial t},\qquad
\partial_1=\frac{\partial}{\partial x},\qquad
\partial_\pm=\frac{1}{2}(\partial_0\pm\partial_1).
\end{equation} 
The Minkowski metric and antisymmetric tensor components are
\begin{equation}
\eta_{00}=-\eta_{11}=\epsilon_{01}=-\epsilon_{10}=1,\qquad\quad
\eta_{01}=\eta_{10}=\epsilon_{00}=\epsilon_{11}=0 ,
\end{equation} 
and in light-cone coordinates
\begin{equation}
\eta^{++}=\eta^{--}=\epsilon^{++}=\epsilon^{--}=0,\qquad\quad
\eta^{+-}=\eta^{-+}=\epsilon^{+-}=-\epsilon^{-+}=2.
\end{equation} 

In Euclidean space we use the coordinates $(x^1,x^2)$, where $x^2=-\ii x_0$
and the complex coordinates $z$, $\bar z$ defined by
\begin{equation}
x^+=\ii (x^2-\ii x^1)=\ii z,\qquad\quad
x^-=\ii (x^2+\ii x^1)=\ii \bar z.
\end{equation} 
So we have
\begin{equation}
-x^2=r^2=(x^1)^2+(x^2)^2=z\bar z,\qquad \quad
\partial=\frac{\partial}{\partial z}=\ii\partial_+,\qquad\quad
\bar\partial=\frac{\partial}{\partial \bar z}=\ii\partial_-.
\end{equation}

\subsection{Energy-momentum tensor}

In the IR part of the paper we use the canonical energy-momentum tensor $T^{\mu\nu}$, 
which is symmetric and conserved: 
\begin{equation}
\partial_-T^{--}+\partial_+T^{+-}=0,\qquad\quad
\partial_+T^{++}+\partial_-T^{-+}=0.
\end{equation} 
Its trace is denoted by
\beq
\Theta=T^\mu_{\,\,\mu}=T^{+-}.
\eeq
The normalization of the canonical EM tensor is fixed by requiring 
that the total momentum operator
\begin{equation}
P_\mu=\int{\rm d}x\,T_{0\mu}(x,t) 
\end{equation} 
acts on any local field $\Phi(x,t)$ according to
\begin{equation}
[P_\mu,\Phi(x,t)]=-\ii\partial_\mu\Phi(x,t).
\end{equation} 

In the UV part we use the CFT normalized Virasoro densities $L(z)$, $\bar L(\bar z)$ with
the usual short distance expansion
\begin{equation}
L(z)L(w)\approx \frac{c}{2}\frac{1}{(z-w)^4}+\frac{2L(w)}{(z-w)^2}
+\frac{\partial L(w)}{z-w},
\end{equation}
where $c$ is the Virasoro central charge.
For any chiral primary field $\Phi(z)$ with conformal weight $\Delta$,
\begin{equation}
L(z)\Phi(w)\approx \frac{\Delta\Phi(w)}{(z-w)^2}+\frac{\partial \Phi(w)}{z-w}.
\end{equation} 
There are analogous formulas for antichiral fields.

The identification of UV and IR fields is given by
\begin{equation}
L=\frac{\pi}{2}\,T^{--}=2\pi T_{++}.
\end{equation} 
Similarly 
\begin{equation}
\bar L=\frac{\pi}{2}\,T^{++}=2\pi T_{--},\qquad\quad 
\tau=\frac{\pi}{2}\,T^{+-}=\frac{\pi}{2}\,\Theta,
\end{equation} 
where $\tau$ is the trace of the EM tensor in CFT normalization.

\subsection{Equal time commutators in CFT}

Equal time commutators are given in the CFT limit by the formulas
\begin{equation}
[P^-,\Phi(z,\bar z)]=-\pi \oint\frac{{\rm d}w}{2\pi \ii}T^{--}(w)\Phi(z,\bar z)  ,
\end{equation} 
\begin{equation}
[P^+,\Phi(z,\bar z)]=-\pi \oint\frac{{\rm d}\bar w}{2\pi \ii}T^{++}(\bar w)\Phi(z,\bar z) .
\end{equation} 
Analogous formulas hold for any chiral conserved currents and charges.

\subsection{Master formula}
\label{subsect.master}

The master formula for the first order conformal perturbation is
\begin{equation} 
\bar\partial L_s(z,\bar z)= 
-\pi\oint\frac{{\rm d}w}{2\pi \ii}{\cal L}_{\rm pert}(w,\bar z)L_s(z)
\end{equation} 
for any  chiral field $L_s(z)$ (in the CFT limit). Applying this to $L(z)$ we obtain,  
for a perturbation by a primary field with conformal weight $\Delta$,
\begin{eqnarray}
\bar\partial L(z,\bar z)&=& 
-\pi\oint\frac{{\rm d}w}{2\pi \ii}{\cal L}_{\rm pert}(w,\bar z)L(z) 
= -\pi\oint\frac{{\rm d}w}{2\pi \ii}\left[\frac{\Delta{\cal L}_{\rm pert}(w,\bar z)}{(z-w)^2}
+\frac{\partial_w{\cal L}_{\rm pert}(w,\bar z)}{z-w}\right]\nonumber \\ 
&=&\pi(1-\Delta)\partial{\cal L}_{\rm pert}(z,\bar z)=-\partial\tau.
\end{eqnarray} 
Thus we conclude that the CFT normalized trace is
\begin{equation}
\tau=-\pi(1-\Delta){\cal L}_{\rm pert},
\end{equation} 
whereas the trace of the canonical EM tensor is
\begin{equation}
\Theta=-2(1-\Delta){\cal L}_{\rm pert}.
\end{equation}

\section{Characters}
\label{app.b}

In this appendix, we summarize the relations among the $su(2)_k$ and the Virasoro characters, 
and the $su(3)_2$ string functions, which are used to confirm the relations among
the coset theories and the minimal models in (\ref{coset}) and (\ref{cosetMM}).

\subsection{$su(2)_k$ and Virasoro characters}

A unitary
highest weight representation of $su(2)_k$ $(k \in \bbZ_{>0})$ has spin 
$l= 0, 1/2,\dots, k/2$, and the
central charge of the corresponding CFT is $c(su(2)_k)= 3k/(k+2)$.
We denote the character of the representation with spin $l$  by
\begin{equation}
\ch_{k,l}(\tau,\theta) := \tr\bigl(q^{L_0-c/24} \ee^{\ii\theta J_0^3} \bigr) \comma
\end{equation}
where $q=\ee^{2\pi \ii \tau}$, $c=c(su(2)_k)$,  and 
 $L_0$ and $J^3_0$ are the zero-modes of the Virasoro generators and 
one of the affine $su(2)$ currents.

The unitary minimal model $\calM_{m,m+1}$ has the central charge
$c(\calM_{m,m+1}) = 1-\frac{6}{m(m+1)} =: c_m $. The spectrum consists of 
the primary  fields $\phi^{(m)}_{r,s}$ with dimensions
\begin{equation}
\label{eq.hrs}
 h_{r,s}^{(m)}  = \frac{[(m+1)r -ms]^2-1}{4m(m+1)} \comma
\end{equation}
where $r=1, \dots, m-1$; $s=1, \dots, r$. By the invariance under $r \to m-r$
and $s\to m+1-s$, the range of $s$ may be extended to $s = 1, \dots, m$.
The character of the representation with $(c_m,h_{r,s}^{(m)})$ is given by
\begin{equation}
  \label{VirasoroCh}
  \chi^{(m)}_{h_{r,s}}(\tau) := \tr\bigl( q^{L_0 - c_m/24}  \bigr) 
  = 
  \eta^{-1}(\tau) \Bigl[ \vartheta_{r(m+1)-sm,m(m+1)}(\tau) - \vartheta_{r(m+1)+sm,m(m+1)}(\tau) \Bigr]
  \comma
\end{equation}
where 
\begin{equation}
   \eta(\tau) = q^{\frac{1}{24}} \prod_{n=1}^\infty (1-q^n) \comma \qquad 
   \vartheta_{m,k}(\tau) = \sum_{n \in \bbZ} q^{k \bigl(n+\frac{m}{2k}\bigr)^2} \period
\end{equation}
The superscript of $h^{(m)}_{r,s}$ has been omitted.

In terms of these characters, the coset representation of the minimal models
$su(2)_{m-2} \times su(2)_{1}/su(2)_{m-1} = \calM_{m,m+1}$  implies 
\cite{Goddard:1986ee}
\begin{equation}
\label{su2decomp}
  \ch_{m-2,l}(\tau,\theta) \, \ch_{1,\epsilon}(\tau,\theta) 
  = \sum_s \ch_{m-1,(s-1)/2}(\tau,\theta)  \chi^{(m)}_{h_{r,s}}(\tau)
  \comma
\end{equation}
where $\epsilon= 0,1/2$; 
$r=2l+1$; $1 \leq r \leq m-1$; $1\leq s \leq m$;
and $r-s $ is even if $\epsilon=0$ and odd if $\epsilon=1/2$.
For $n=3$ the relation (\ref{cosetMM}) reads 
\begin{equation}
  \frac{su(3)_2}{u(1)^2} \cong 
  \frac{su(2)_1 \times su(2)_1}{su(2)_2} \times \frac{su(2)_2 \times su(2)_1}{su(2)_3}
  \comma
\end{equation}
with the two factors $su(2)_2$ being identified.
By this identification, 
 the coset partition function consists of the terms of the form  $\chi^{(3)}_h \chi^{(4)}_{h'}$ 
 where the two Virasoro characters
share the common $su(2)_2$ in the decompositions,
$\ch_{1,l} \, \ch_{1,\epsilon} = \sum_s \ch_{2,(s-1)/2} \chi^{(3)}_h$ and
$\ch_{2,l} \, \ch_{1,\epsilon} = \sum_s \ch_{3,(s-1)/2} \chi^{(4)}_h$.
For example, one has $\chi^{(3)}_{h_{r,1}}$ $(r=1,2)$ 
with $\ch_{2,0}$ on the right side of the first decomposition, and 
$\chi^{(4)}_{h_{1,s}}$ $(s=1, \dots, 4)$ with $\ch_{2,0}$ on the left side 
of the second decomposition. 
 This gives $ \phi_{r,1}^{(3)} \times  \phi_{1,s}^{(4)}  $  $(r=1,2; s=1, ..., 4)$ in the spectrum,
which have the form of the projected products $ \phi_{r,p}^{(m)} \phi_{p,s}^{(m+1)}$ \cite{Crnkovic:1989ug}.
Taking into account  $\phi_{r,s}^{(m)} = \phi_{m-r, m+1-s}^{(m)}$, 
and reducing the multiplicities by a factor two so that the identity appears only once, 
one finds the spectrum of $su(3)_2/u(1)^2$ in terms of the primaries of $\calM_{3,4}$
and $\calM_{4,5}$ as in (\ref{FieldContent}).

\subsection{$su(3)_2$ string functions and Virasoro characters in 
$\calM_{3,4}, \calM_{4,5}$}

Chiral fields in the $g_k/u(1)^{r_g}$ coset 
(generalized parafermion)  theory are labeled by the highest weight $\Lambda$
and the weight $\lambda$ of $g_k$ as $\Phi^\Lambda_\lambda$.
The parafermionic character for  $\Phi^\Lambda_\lambda$ is 
written as $\ch^\Lambda_\lambda(\tau) := \tr (q^{L_0 -c/24}) 
= \eta(\tau)^{r_g} c^\Lambda_\lambda(\tau)$,
where $c$ is the central charge, $r_g$ is the rank of $g$ 
and $c^\Lambda_\lambda$ is the string function.
 
For $su(3)_2$, there are four independent string functions. Using the Dynkin labels,
they read \cite{Kac:1984mq}
\begin{eqnarray}
  c^{110}_{110}(\tau) & = & \eta(\tau)^{-4}\eta(2\tau) q^{1/20} 
   \prod_{n \in \bbZ_{>0}, n \neq \pm 1({\rm mod \, 5})} (1-q^{2n}) \comma \nn \\
   c^{200}_{011}(\tau) & = & \eta(\tau)^{-4}\eta(2\tau) q^{9/20} 
   \prod_{n \in \bbZ_{>0}, n \neq \pm 2({\rm mod \, 5})} (1-q^{2n}) \comma \nn \\
   c^{200}_{200}(\tau) - c^{200}_{011}(\tau) & = &   \eta(\tau)^{-4}\eta(\tau/2) q^{1/80} 
   \prod_{n \in \bbZ_{>0}, n \neq \pm 1({\rm mod \, 5})} (1-q^{n/2}) \comma \\
   c^{110}_{110}(\tau) - c^{110}_{002}(\tau) & = &  \eta(\tau)^{-4}\eta(\tau/2) q^{9/80} 
   \prod_{n \in \bbZ_{>0}, n \neq \pm 2({\rm mod \, 5})} (1-q^{n/2}) \period \nn 
\end{eqnarray}
These are related to the products of the Virasoro characters in $\calM_{3,4} \times \calM_{4,5}$
as \cite{Ninomiya:1986dp}
\begin{eqnarray}
\label{CchV}
\eta(\tau)^2 c^{110}_{110}(\tau) & = &
 \chi_{\frac{1}{16} \frac{3}{80}}(\tau)  = q^{-\frac{c}{24} +\frac{1}{10}} + \cdots \comma \nn \\
 \eta(\tau)^2 c^{200}_{011}(\tau) & = & 
 \chi_{\frac{1}{16} \frac{7}{16}}(\tau) =  q^{-\frac{c}{24} +\frac{1}{2}} + \cdots \comma \nn \\
  \eta(\tau)^2 c^{200}_{200}(\tau) & = & 
 \chi_{0 0}(\tau)  +  \chi_{\frac{1}{2} \frac{3}{2}}(\tau) = q^{-\frac{c}{24} +0} + \cdots \comma  \\
  \eta(\tau)^2 c^{110}_{002}(\tau) & = & 
 \chi_{0 \frac{3}{5}}(\tau)  +  \chi_{\frac{1}{2} \frac{1}{10}}(\tau)
 = 2 q^{-\frac{c}{24} +\frac{3}{5}} + \cdots \comma \nn
\end{eqnarray}
where $c = c_3 + c_4 = 6/5$ and 
\begin{equation}
     \chi_{h h'} (\tau) := 
     \chi^{(3)}_{h}(\tau) \times 
     \chi^{(4)}_{h'}(\tau) \period
\end{equation}
In the main text, we have denoted 
$ \chi_h^{(m)}$ $(m=3,4)$ by $\chi_h^{(i)}$ $(i=1,2)$.
Furthermore, with (\ref{VirasoroCh}) one can check that 
\begin{equation}
     \chi_{\frac{1}{16} \frac{3}{80}}(\tau) 
    =  \chi_{0 \frac{1}{10}}(\tau) + \chi_{\frac{1}{2} \frac{3}{5}}(\tau)
     \comma \qquad
  \chi_{\frac{1}{16} \frac{7}{16}}(\tau) 
  = \chi_{\frac{1}{2}  0}(\tau) + \chi_{0  \frac{3}{2}}(\tau) \period
\end{equation}

Since the modular invariant for $su(3)_2$ is unique and diagonal
\cite{Gannon:1992ty}, so is the modular invariant 
for $su(3)_2/u(1)^2$: 
\bea
 Z\bigl(su(3)_2/u(1)^2\bigr)
  =  \sum \bigl{|} \ch^\Lambda_\lambda (\tau) \bigr{|}^2 && \nn \\
  &&\hspace{-5cm} = \bigl{|} \eta(\tau)\bigr{|}^4 \Bigl(
      \bigl{|} c ^{200}_{200} (\tau) \bigr{|}^2 + 3 \bigl{|} c ^{200}_{011} (\tau) \bigr{|}^2
      + 
       3\bigl{|} c ^{110}_{110} (\tau) \bigr{|}^2 + \bigl{|} c ^{110}_{002} (\tau) \bigr{|}^2
       \Bigr) \period
\eea
From the relations among the string functions and the Virasoro characters given above,
one finds that this modular invariant agrees with the one in (\ref{eq.cosetchara}). 
Given the multiplicities which are read off
from the rightmost expressions in (\ref{CchV}), one confirms the chiral field content:
1 identity, 3 fields with $h=1/2$, 3 fields with $h=1/10$ and 2 fields with $h=3/5$.

\section{Conserved charges from the counting argument}
\label{app.count}

In this appendix we analyze conserved charges in the product picture in Section \ref{subsec.MMrep}. 

\paragraph{Spin 1 charges}

Let us see how the counting argument works for the spin $s=2$ currents. We focus on the left chiral dependence
as the right chiral part behaves as a spectator. We have 3 candidates
to remain conserved after the perturbation, which correspond to the
vectors\footnote{Using the state-operator correspondence we often represent field operators by 
their corresponding vectors.}: 
\begin{equation}
\vert L_{-2}^{(1)}\rangle=\frac{1}{2}\vert\psi_{-\frac{3}{2}}\psi_{-\frac{1}{2}}\rangle
 \, , \qquad 
\vert L_{-2}^{(2)}\rangle\, ,  \qquad 
\vert L_{-2}^{(3)}\rangle=\vert\psi_{-\frac{1}{2}}G_{-\frac{3}{2}}\rangle
\, .
\end{equation}
These are the holomorphic stress tensor components in each theory
$L^{(i)}(z)$ and the product $L^{(3)}(z)=\psi(z)G(z)$. Clearly none is a total derivative.
After the perturbation
the level 1 subspace contains 3 vectors: $\psi_{-\frac{3}{2}}\vert{\scriptstyle \frac{1}{10}}\rangle$,
$L_{-1}G_{-\frac{1}{2}}\vert{\scriptstyle \frac{1}{10}}\rangle$ and
$L_{-1}\psi_{-\frac{1}{2}}\vert{\scriptstyle \frac{1}{10}}\rangle$,
out of which only 1 is not a total derivative. The two total derivatives
are the descendants of $\psi_{-\frac{1}{2}}\vert{\scriptstyle \frac{1}{10}}\rangle \sim 
\vert \Phi_{1j}\rangle $
and $G_{-\frac{1}{2}}\vert{\scriptstyle \frac{1}{10}}\rangle \sim \vert 
\Phi_{2j} \rangle $ as we are focusing
only on the left chiral dependence. Comparing
the dimensions we can conclude that two appropriate linear combinations of 
the $L^{(i)}$ have to be conserved. Clearly one of them corresponds
to the energy $L(z)=L^{(1)}(z)+L^{(2)}(z)$. The existence of the other conserved 
charge is consistent with the finding from the IR side
(see Section \ref{sec.AnMCrel}) and can be obtained from short distance OPEs. 

We calculate the relevant terms one by one:
\begin{equation}
L_{0}^{(1)}\vert\Phi_{1j}\rangle=\frac{1}{2}\vert\Phi_{1j}\rangle \, , \qquad 
L_{-1}^{(1)}\vert\Phi_{1j}\rangle=L_{-1}\vert\Phi_{1j}\rangle
    -\psi_{-\frac{1}{2}}\bar{\psi}_{-\frac{1}{2}}^{(j)}L_{-1}\vert\Phi\rangle \, , 
\end{equation}
\begin{equation}
\qquad L_{n}^{(1)}\vert\Phi_{2j}\rangle=0  
\quad (n=0,-1) \, .
\end{equation}
The action of $L^{(2)}$ is 
\begin{equation}
L_{0}^{(2)}\vert\Phi_{1j}\rangle=\frac{1}{10}\vert\Phi_{1j}\rangle \, , \quad 
L_{0}^{(2)}\vert\Phi_{2j}\rangle=\frac{3}{5}\vert\Phi_{2j}\rangle \, , \quad 
L_{-1}^{(2)}\vert\Phi_{ij}\rangle=
\psi_{-\frac{1}{2}}^{(i)}\bar{\psi}_{-\frac{1}{2}}^{(j)}L_{-1}\vert\Phi\rangle \, . 
\end{equation}
Finally the action of $L_{0}^{(3)}=\dots\psi_{-\frac{1}{2}}G_{\frac{1}{2}}+\psi_{\frac{1}{2}}G_{-\frac{1}{2}}+\dots$
and $L_{-1}^{(3)}=\dots\psi_{-\frac{3}{2}}G_{\frac{1}{2}}
+\psi_{-\frac{1}{2}}G_{-\frac{1}{2}}+\psi_{\frac{1}{2}}G_{-\frac{3}{2}}+\dots$
turns out to be 
\begin{equation}
L_{0}^{(3)}\vert\Phi_{1j}\rangle=\frac{1}{\sqrt{5}}\vert\Phi_{2j}\rangle \, , \qquad 
L_{-1}^{(3)}\vert\Phi_{1j}\rangle=\frac{\sqrt{5}}{3}L_{-1}\vert\Phi_{2j}\rangle \, , 
\end{equation}
\begin{equation}
L_{0}^{(3)}\vert\Phi_{2j}\rangle=\frac{1}{\sqrt{5}}\vert\Phi_{1j}\rangle \, , \qquad 
L_{-1}^{(3)}\vert\Phi_{2j}\rangle=\frac{1}{\sqrt{5}}L_{-1}\vert\Phi_{1j}\rangle+\frac{4}{\sqrt{5}}
 \psi_{-\frac{1}{2}}^{(1)}\bar{\psi}_{-\frac{1}{2}}^{(j)}L_{-1}\vert\Phi\rangle \, , 
\end{equation}
where we used the super null vector 
$G_{-\frac{3}{2}}\vert\Phi\rangle=\frac{5}{3}G_{-\frac{1}{2}}L_{-1}^{(2)}\vert\Phi\rangle$
of the superconformal algebra. These formulas are used to calculate explicitly the second spin $1$ charge
in Subsection \ref{cons.charge}.

\paragraph{Spin $2$ charges}

In order to prove the factorization of the scattering matrix we need
at least one higher spin charge. In \cite{FernandezPousa:1997zb}  
the authors used the coset chiral algebra and showed the existence
of spin 2 conserved charges. As we are working with a smaller chiral
algebra the counting argument does not guarantee any conserved charge at this 
level.
Indeed, the possible candidates at the third level are
\begin{equation}
\vert\psi_{-\frac{5}{2}}\psi_{-\frac{1}{2}}\rangle \, , \quad 
\vert L_{-3}^{(2)}\rangle \, , \quad 
\vert G_{-\frac{5}{2}}\psi_{-\frac{1}{2}}\rangle \, , \quad 
\vert G_{-\frac{3}{2}}\psi_{-\frac{3}{2}}\rangle \,  ,
\label{level3}
\end{equation}
out of which only one is not a total derivative.
On the other hand after the perturbation the level $2$ descendant space is 
\begin{equation}
L_{-2}^{(2)}G_{-\frac{1}{2}}\vert{\scriptstyle \frac{1}{10}}\rangle \, , \quad 
(L_{-1}^{(2)})^{2}G_{-\frac{1}{2}}\vert{\scriptstyle \frac{1}{10}}\rangle \, , \quad 
(L_{-1}^{(2)})^{2}\psi_{-\frac{1}{2}}\vert{\scriptstyle \frac{1}{10}}\rangle \, , \quad 
G_{-\frac{1}{2}}\psi_{-\frac{3}{2}}\psi_{-\frac{1}{2}}\vert{\scriptstyle \frac{1}{10}}\rangle \, ,
\label{eq.level21}
\end{equation}
\begin{equation}
L_{-1}^{(2)}\psi_{-\frac{3}{2}}\vert{\scriptstyle \frac{1}{10}}\rangle \ , \qquad 
\psi_{-\frac{5}{2}}\vert{\scriptstyle \frac{1}{10}}\rangle \, , 
\label{eq.level22}
\end{equation}
which contains three non-derivative operators and does not guarantee the existence
of any conserved charge at this level.
The reason why we could not find the spin 2 conserved charges is that we did not include
in our chiral space (\ref{level3}) the contributions of the other two fermions 
of the representation spaces $2\chi_{\frac{1}{16}\frac{7}{16}}
\bar{\chi}_{\frac{1}{16}\frac{7}{16}}$.

\paragraph{Spin $3$ charges and integrability}

Contrary to the spin 2 case our chiral algebra will be sufficient 
to find conserved charges at spin 3.
In this case, we first analyze the operators of the chiral
algebra $\mathcal{A}$ at level 4. We list the corresponding vectors:
\begin{equation}
\vert\psi_{-\frac{7}{2}}\psi_{-\frac{1}{2}}\rangle \, , \quad 
\vert\psi_{-\frac{5}{2}}\psi_{-\frac{3}{2}}\rangle \, ,  \quad 
\vert L_{-2}^{(2)}\psi_{-\frac{3}{2}}\psi_{-\frac{1}{2}}\rangle \, , \quad 
\vert L_{-2}^{(2)}L_{-2}^{(2)}\rangle \ , \quad 
\vert L_{-4}^{(2)}\rangle \ ,
\end{equation}
\begin{equation}
\vert L_{-2}^{(2)}G_{-\frac{3}{2}}\psi_{-\frac{1}{2}}\rangle \, , \quad 
\vert G_{-\frac{7}{2}}\psi_{-\frac{1}{2}}\rangle \, , \quad 
\vert G_{-\frac{5}{2}}\psi_{-\frac{3}{2}}\rangle \, , \quad 
\vert G_{-\frac{3}{2}}\psi_{-\frac{5}{2}}\rangle \, .
\end{equation}
To see how many of them is not a total derivative we compare them to  
the states at the third level (\ref{level3}) and conclude that we have 5 non-derivative
operators.
As for the subspace after the perturbation, 
at the level 3 it contains the operators,
\begin{equation}
(L_{-1}^{(2)})^{3}\psi_{-\frac{1}{2}}\vert{\scriptstyle \frac{1}{10}}\rangle \, , \quad 
L_{-1}^{(2)}L_{-2}^{(2)}\psi_{-\frac{1}{2}}\vert{\scriptstyle \frac{1}{10}}\rangle \, , \quad 
L_{-1}^{(2)}L_{-2}^{(2)}G_{-\frac{1}{2}}\vert{\scriptstyle \frac{1}{10}}\rangle \, , \quad 
(L_{-1}^{(2)})^{3}G_{-\frac{1}{2}}\vert{\scriptstyle \frac{1}{10}}\rangle \, ,
\end{equation}
\begin{equation}
(L_{-1}^{(2)})^{2}\psi_{-\frac{3}{2}}\vert{\scriptstyle \frac{1}{10}}\rangle  \, , \quad 
L_{-1}^{(2)}G_{-\frac{1}{2}}\psi_{-\frac{3}{2}}
\psi_{-\frac{1}{2}}\vert{\scriptstyle \frac{1}{10}}\rangle \, , \quad 
L_{-1}^{(2)}\psi_{-\frac{5}{2}}\vert{\scriptstyle \frac{1}{10}}\rangle \, , \quad 
G_{-\frac{1}{2}}\psi_{-\frac{5}{2}}\psi_{-\frac{1}{2}}\vert{\scriptstyle \frac{1}{10}}\rangle \, , \quad 
\psi_{-\frac{7}{2}}\vert{\scriptstyle \frac{1}{10}}\rangle \, , 
\end{equation}
Again to see how many of them is not a total derivative we recall  the
states at one level higher (\ref{eq.level21}), (\ref{eq.level22}). 
Thus we have  three non-derivative operators. This means that we can make
 two spin 3 conserved charges.
This assures the quantum integrability of this model, as shown in \cite{FernandezPousa:1997zb}. 
Clearly the compatibility of the perturbations $\Phi_{i1}$ and $\Phi_{i2}$ again
forces the coupling constant to factorize $\nu_{12}\nu_{21}=\nu_{11}\nu_{22}$.

\section{Projected tensor product of minimal models}
\label{app.proj}

With extension to general  cases in mind, in this appendix
we discuss the identification between the  $su(n)_2/u(1)^{n-1} $ coset CFT and 
the projected tensor product of the minimal models
\begin{align}
{su(n)_2\over u(1)^{n-1}} =
\mathbb{P}({\cal M}_{3,4}\times \cdots \times {\cal M}_{n+1,n+2}).
\label{eq:coset1}
\end{align}
In the coset model $su(n)_2/u(1)^{n-1}$, there are $n-1$ weight zero primary fields in the adjoint
representation of $su(n)$, whose conformal dimension is ${n\over n+2}$, and which are used as 
the perturbation operators.
In the projected tensor product the corresponding operators are 
represented as 
\begin{align}
 \prod_{m=3}^{n+1}\phi^{(m)}_{k_m, k_{m+1}} ,
 \label{eq:projp2}
\end{align}
where $ k_m=1,3$ with $k_m \leq k_{m+1}$, $k_{n+1}=3$  \cite{Kac:1988tf}.
Here, the degenerate primary fields $\phi^{(m)}_{r,s}$ have conformal dimension
$h_{r,s}^{(m)}$ as in  (\ref{eq.hrs}). 
Since $k_m$ change only once, the products are of the form 
$1 \times \cdots 1 \times \phi_{1,3}^{(p)} \times  \phi_{3,3}^{(p+1)} \times \cdots  \phi_{3,3}^{(n+1)}$
($ 3 \leq p \leq n+1$).
Their conformal dimension is shown to be
\begin{align}
\sum_{m=3}^{n+1}h^{(m)}_{k_m,k_{m+1}}
=\frac{p-1}{p+1}+\sum_{m=p+1}^{n+1}\frac{2}{m(m+1)}={n\over n+2} .
\end{align}
For example, in the $su(3)_2/u(1)^2$ model, one has the primary fields 
$\phi^{(3)}_{1,1}\phi^{(4)}_{1,3}$ and $\phi^{(3)}_{1,3}\phi^{(4)}_{3,3}$ in
the projected product ${\mathbb P}({\cal M}_{3,4}\times {\cal M}_{4,5})$ as explained 
in Section \ref{sec.HSGpCFT}.

In order to show the integrability of the HSG model, it is necessary to construct the conserved 
currents with integer spins.
The quantum conserved currents with spin two and three have been constructed in \cite{FernandezPousa:1997zb}. 
In the projected product of minimal models, 
candidates of spin two conserved currents consist of 
the energy-momentum currents $L^{(m)}(z)$ for each minimal model ${\cal M}_{m,m+1}$,
and spin two operators
$\phi^{(m)}_{1,3}\phi^{(m+1)}_{3,1}$.
They thus take the form
\begin{align}
\Lambda=\sum_{l=3}^{n+1}\alpha_l L^{(l)}+\sum_{m=1}^{n}\beta_m  
\phi^{(m)}_{1,3}\phi^{(m+1)}_{3,1}
\end{align}
with some coefficients $\alpha_l$ and $\beta_m$.

For the $su(3)_2/u(1)^2$ model, the primary field $ \phi^{(3)}_{1,3}\phi^{(4)}_{3,1} $
is identified with the $L^{(i=3)}$ in the notation in Section \ref{sec.HSGpCFT} and 
Appendix \ref{app.count}.
Focusing on each chiral sector, the projected product can be reorganized
into an ordinary tensor product in this special case \cite{Crnkovic:1989ug},
where the Virasoro characters $\chi_h^{(m)}$ $(m=3,4)$ are linearly combined 
into the chiral characters of the free fermion and the ${\cal N}=1$ super minimal model, respectively.

\section{Form factors}
\label{sect.formfactor}

In this appendix we give all higher form factors corresponding to our tensor operators. 
We adapted the results of \cite{CastroAlvaredo:2000em,CastroAlvaredo:2000nk} to our 
form factor conventions and field normalizations.

The $n$-particle form factors of a local field operator $X$ are defined by the 
matrix elements 
\beq
{\cal F}^X_{a_1\dots a_n}(\theta_1,\dots,\theta_n)=\langle 0\vert X(0)\vert
\theta_1,a_1;\dots;\theta_n,a_n\rangle,
\eeq
where particle states are normalized according to
\beq
\langle \theta^\prime,a^\prime\vert\theta,a\rangle=\delta_{a,a^\prime}
\delta(\theta-\theta^\prime).
\eeq
Below we give the \lq\lq scalarized'' form factors for our tensor operators for the
case of $\ell=2s$ type-1 particles ($s\geq1$) and $m=2t$ type-2 particles ($t\geq1$). 
The total particle number
is $n=\ell+m$. The form factor polynomial can be written
\beq
q_{a_1\dots a_n}(x_1,\dots,x_n)=\widetilde H^{\ell, m}\,
\widetilde Q^{\ell,m}_{a_1\dots a_n}(x_1,\dots,x_n), 
\eeq
where the normalization constant $\widetilde H^{\ell, m}$ is given by
\beq
\widetilde H^{\ell, m}=\frac{(4\pi\ii)^{s(\ell-1)}}{(2\pi)^s}\,
\widetilde H^{0, m}.
\eeq
The lowest constants $\widetilde H^{0, m}$ still must be fixed from some further
considerations. For example, from the normalization of the 2-particle form factors we can 
determine
\beq
\widetilde H^{0,2}=\ii.
\eeq
The polynomials are given as
\beq
\widetilde Q^{\ell, m}=(-1)^{(s+1)t}{\rm e}^{-t\sigma}\left(\Sigma^{(1)}\right)^{s-t}
\left(\Sigma^{(2)}\right)^{t}\,{\cal D}^{s,t},
\eeq
where
\beq
\Sigma^{(a)}=\prod_{a_i=a}x_i
\eeq
and ${\cal D}^{s,t}$ is the determinant of an $(s+t-2)\times (s+t-2)$ matrix,
\beq
{\cal D}^{1,1}=1,\qquad\quad {\cal D}^{s,t}=\det({\cal M}^{s,t}),
\eeq
whose matrix elements are symmetric polynomials,
\beq
\left({\cal M}^{s,t}\right)_{ij}=\left\{
\begin{split}
\sigma^{(1)}_{2j-2i+1}\qquad\qquad\quad&1\leq i<t,\\
(-1)^{j-i+t}\hat\sigma^{(2)}_{2j-2i+2t-1}\qquad &t\leq i\leq s+t-2.
\end{split}\right.
\eeq
The symmetric polynomials are defined by
\beq
\prod_{a_i=1}(z+x_i)=\sum_{k=-\infty}^\infty z^{\ell-k}\sigma^{(1)}_k,\qquad\quad
\prod_{a_i=2}(z+x_i{\rm e}^{-\sigma})=
\sum_{k=-\infty}^\infty z^{m-k}\hat\sigma^{(2)}_k.
\eeq
Special cases are
\beq
\sigma^{(1)}_1=\hat P_{(1)}^+,\qquad
\hat\sigma^{(2)}_1={\rm e}^{-\sigma}\hat P_{(2)}^+,\qquad
\sigma^{(1)}_\ell=\Sigma^{(1)},\qquad
\hat\sigma^{(2)}_m={\rm e}^{-m\sigma}\Sigma^{(2)}.
\eeq

\section{Generalized $\Theta$ sum rule}
\label{app.d}

In this appendix we describe a generalization of the well-known $\Theta$ 
sum rule \cite{DSC}, 
which is used in Section \ref{sec.AnMCrel}.
Let us consider a conserved spin-2 current $Y^{\mu\nu}$:
\begin{equation}
\partial_\mu Y^{\mu\nu}=0. 
\end{equation} 
We do not assume that $Y^{\mu\nu}$ is symmetric and it need not be conserved in its second tensor
index. Moreover, we do not assume that the theory is parity invariant.

Let us consider the Euclidean 2-point correlation function
\begin{equation}
C^{\mu\nu}(x)=\langle Y^{\mu\nu}(x)\Psi(0)\rangle_c,
\end{equation} 
where $\Psi$ is some scalar field. From Euclidean (Lorentz)
covariance it must be of the form
\begin{equation}
C^{\mu\nu}(x)=-x^\mu x^\nu\frac{F(r^2)}{r^4}+\eta^{\mu\nu}\frac{A(r^2)}{r^2}
+\epsilon^{\mu\nu}\frac{B(r^2)}{r^2} ,
\end{equation} 
and its components are
\begin{equation}
C^{+-}=\frac{F(r^2)+2A(r^2)+2B(r^2)}{r^2}=\frac{G(r^2)}{r^2},\qquad\quad
C^{--}=\frac{\bar z^2F(r^2)}{r^4}=\frac{F(r^2)}{z^2}.
\end{equation} 
The conservation equation
\begin{equation}
\partial C^{+-}+\bar\partial C^{--}=0
\end{equation} 
is equivalent to
\begin{equation}
\frac{G}{r^2}=(F+G)^\prime,
\end{equation} 
where $^\prime$ here means derivative with respect to the argument $r^2$. From here we have
\begin{equation}
\int {\rm d}^2x\langle Y^{+-}(x)\Psi(0)\rangle_c=\pi\int_0^\infty{\rm d}r^2\frac{G(r^2)}{r^2}=
\pi(F+G)\vert_0^\infty.
\end{equation} 
We assume that the theory is massive and therefore
\begin{equation}
F(\infty)=G(\infty)=0.
\end{equation} 
We also assume that the relevant conformal weights are $\Delta<1$ and so
\begin{equation}
G(0)=0.
\end{equation} 
We conclude that the integral of the scalar component is completely determined by the
short distance asymptotics of the tensor component:
\begin{equation}
\int {\rm d}^2x\langle Y^{+-}(x)\Psi(0)\rangle_c=-\pi F(0),
\end{equation} 
where
\begin{equation}
\langle Y^{--}(x)\Psi(0)\rangle\approx \frac{F(0)}{z^2}.
\end{equation} 
If we apply these formulas to the EM tensor $T^{\mu\nu}$ and
$\Psi$ is a scalar field with conformal weight $\Delta$, we have
\begin{equation}
F(0)=\frac{2\Delta\langle\Psi\rangle}{\pi}
\end{equation} 
and
\begin{equation}
\int {\rm d}^2x\langle \Theta(x)\Psi(0)\rangle_c=-2\Delta\langle\Psi\rangle.
\end{equation} 
For the CFT normalized trace we have
\begin{equation}
\int {\rm d}^2x\langle \tau(x)\Psi(0)\rangle_c=-\pi\Delta\langle\Psi\rangle.
\end{equation} 
This is the $\Theta$ sum rule in its original form \cite{DSC}.

\section{Symmetries of the mass-coupling relation}
\label{app.e}

In this appendix, we describe symmetries of the mass functions $\mu_a(\lambda_1,\lambda_2)$,
and their parametrization invariant under the symmetries.

\subsection{$S_3$ Weyl symmetry}

Our $\mu_a(\lambda_1,\lambda_2)$ functions ($a=1,2$) satisfy the differential 
equation (\ref{eq.diff}), as well as the scaling equation,
\beq
  {\cal L} \mu_a = \frac{5}{2} \mu_a \comma \qquad 
  {\cal L} = \lambda_1 \del_1 + \lambda_2 \del_2 \period
\eeq
The transformation rules for $\mu_a$ under the $S_3$ Weyl symmetry
are
\beq
  \hat\mu_a(\lambda_1,\lambda_2) = \mu_a (-\lambda_1, \lambda_2)
  \comma \qquad
  \check\mu_a(\lambda_1,\lambda_2) = \mu_a (\check\lambda_1, \check\lambda_2)
  \comma
\eeq
where 
\beq
  \check\lambda_1 = -\frac{1}{2} \lambda_1 + \frac{\sqrt{3}}{2} \lambda_2
  \comma \qquad
  \check\lambda_2 = -\frac{\sqrt{3}}{2} \lambda_1 - \frac{1}{2} \lambda_2 \comma
\eeq
corresponding to a clockwise rotation by 120 degrees.
Our differential equations are consistent with these discrete symmetries,
since we can show that $\hat\mu_a$ and $\check\mu_a$ satisfy the same equations
 as $\mu_a$.
Using these one can extend the solution (\ref{eq.mulam}) outside the fundamental domain.

\subsection{$\mu_1 \Leftrightarrow \mu_2$  chiral Dynkin reflection}

Next, let us consider the transformation,
\beq
   \tilde\lambda_1 = -\frac{1}{2} \lambda_1 + \frac{\sqrt{3}}{2} \lambda_2
   \comma \qquad 
  \tilde \lambda_2 = \frac{\sqrt{3}}{2} \lambda_1 + \frac{1}{2} \lambda_2 \comma
\eeq
which is the reflection with respect to the axis $ \lambda_2 = \sqrt{3} \lambda_1$
($\eta = 1/\sqrt{3}$). Using the explicit solution in Subsection \ref{subsec.solDE},
we can show that 
\beq
  \mu_1(\tilde\lambda_1,\tilde\lambda_2) = \mu_2(\lambda_1,\lambda_2)
  \period
\eeq
From this symmetry it is sufficient to consider half of the fundamental domain,
given by $0 \leq \eta \leq 1/\sqrt{3}$. The other half $1/\sqrt{3} \leq \eta \leq \sqrt{3}$
is mapped to the first half by this  $\mu_1 \Leftrightarrow \mu_2$ symmetry.

\subsection{$S_3$-invariant parametrization}

To find the expression of $\mu_a(\lambda_i)$ in the entire $(\lambda_1,\lambda_2)$-plane,
it is useful to adopt a $S_3$-invariant parametrization,
\beq
   2 \mu_a = (p_2)^{5/4} f_a(y) \comma
\eeq
where $y= p_3^2/p_2^3$ and $p_2=\lambda_1^2+\lambda_2^2$, 
$p_3 = \lambda_2(\lambda_2^2 - 3 \lambda_1^2)$ as in (\ref{eq.p2}).
We note  $0 \leq y \leq 1$ for real $\lambda_i$.
The differential equation (\ref{eq.diff}) then becomes
\beq
  144y(1-y) f''_a(y) + 72(1-y) f'_a(y) - 5 f_a(y) = 0 \comma
\eeq
whose general solutions are
\beq
   f_a(y) = 
   C_{a1} \cdot {}_2F_1\Bigl(-\frac{5}{12}, -\frac{1}{12}; \frac{1}{2} ; y \Bigr)
  + C_{a2} \cdot \sqrt{y}  \, {}_2F_1\Bigl(\frac{1}{12}, \frac{5}{12}; \frac{3}{2} ; y \Bigr) 
  \period
\eeq
The constants $C_{ak}$ $(a,k=1,2)$ are determined so as to match the mass-coupling
relations in the equal- and single-mass cases discussed in Section \ref{subsec.special},
as done in  \cite{Hatsuda:2011ke}. The results are
\beq
   C_{11} = C_{21} = \tilde{K} \comma \qquad
   C_{22} = - C_{12} = 2 \tilde{K} \frac{\Gamma(\frac{13}{12})\Gamma(\frac{17}{12})}{
   \Gamma(\frac{7}{12})\Gamma(\frac{11}{12})} \comma
\eeq
where $\tilde{K}$ is given in (\ref{eq.Ktilde}). A useful identity in deriving these is
\beq
   2 = \frac{3^{1/4}}{\sqrt{2} \pi^2} \Gamma\Bigl(\frac{1}{4}\Bigr)^{2} 
    \Gamma\Bigl(\frac{7}{12}\Bigr)  \Gamma\Bigl(\frac{11}{12}\Bigr) \period
\eeq
In this expression, $ f_1(y) \leq f_2(y)$ ($0 \leq y \leq 1$). Thus, 
smooth functions $\mu_a(\lambda_i)$ are obtained by 
continuing $f_1$ and $f_2$ along the locus of $f_1(y) = f_2(y)$, where
$y=0$, and $\lambda_2 = \pm \sqrt{3} \lambda_1$ or $\lambda_2 =0$.

\section{$\xi$-$\eta$ relation}
\label{app.f}

In this appendix, we describe the relation between the ratios of the chiral masses $\mu_a$ 
and the couplings $\lambda_i$.
The result is used in Section \ref{sec.NumMCrel}.

For numerical studies we need the $\xi$-$\eta$ relation, where
\beq
   \xi = \frac{\mu_1}{\mu_2} = \frac{q_1(\eta)}{q_2(\eta)} \comma 
   \qquad
   \eta = \frac{\lambda_1}{\lambda_2} \comma
\eeq
and $q_a$ is defined in (\ref{eq.muq}).
The derivative of the $\xi(\eta)$ function is
\beq
   \frac{d\xi}{d\eta} = \frac{W}{q_2^2} \comma
\eeq
where $W$ is the Wronskian of the differential equation:
\beq
  W = q'_1 q_2 - q_1 q'_2 \period
\eeq
It satisfies the differential equation,
\beq
   \frac{W'}{W} = \frac{12 - 2 \eta^2}{ \eta(\eta^2 -3)} \period
\eeq
This can be used to determine $W$ explicitly. We find
\beq
   W(\eta) = \frac{2\sqrt{3} B^2 F(1)}{\eta^4} (3-\eta^2) \comma
\eeq
and
\beq
   \frac{d\xi}{d\eta} =
   \frac{8\sqrt{3}F(1)\eta}
    {(\sqrt{3}-\eta)^3F^2\Bigl(\frac{\sqrt{3}-\eta}{\sqrt{3}+\eta}\Bigr)}
   \comma
\eeq
with $F(z)$ defined in (\ref{eq.Fz}).
From this expression we see that the derivative is positive. This means that
$\xi(\eta)$ is monotonically increasing and its inverse is well-defined. 
In the fundamental domain, $0 \leq \eta \leq \sqrt{3}$, it is
sufficient to determine this inverse function $\eta(\xi)$ for $ 0 \leq \xi \leq 1$ 
($ 0 \leq \eta \leq 1/\sqrt{3}$). For $\xi \geq 1$ ($ 1/\sqrt{3} \leq \eta \leq \sqrt{3}$),
it can be obtained using the formula
\beq
\eta(\xi) = \bar\eta(1/\xi) \comma
\eeq
where
\beq
\bar\eta = \frac{\sqrt{3}-\eta}{1+\sqrt{3}\eta} \period
\eeq

\end{document}